\documentclass[11pt, letter]{article}
\usepackage{graphicx} 

\usepackage[left=1in, right=1in, top=1in, bottom=1in, footskip=.5in]{geometry}
\usepackage[bottom]{footmisc} 
\usepackage{setspace}
\usepackage[caption=false, position=top, labelformat=empty]{subfig}
\usepackage{amsmath,amsfonts,array,graphicx,colortbl,multirow,hhline,calc,tabularx,threeparttable,wrapfig,adjustbox,lscape,booktabs,siunitx,ragged2e}

\usepackage[%
    natbib=true,
    style=apa,
    citestyle=authoryear-comp, 
    backend=biber,
    bibencoding=utf8,
    uniquename=false, 
    maxbibnames=99, 
    nohashothers=true,
    minnames=1, 
    maxnames=3, 
    uniquelist=false 
]{biblatex} 
\usepackage{etoolbox} 
\usepackage{appendix}

\AtBeginEnvironment{appendices}{
    \counterwithin{figure}{section}
    \counterwithin{table}{section}
    \numberwithin{equation}{section}
}
\AtBeginEnvironment{table}{
    \setstretch{1}
}
\AtBeginEnvironment{figure}{
    \setstretch{1}
}

\usepackage{bookmark,hyperref} 
\hypersetup{
    colorlinks   = true,        
    urlcolor     = blue,         
    linkcolor    = blue,         
    citecolor    = blue         
}

\title{StockGPT: A GenAI Model for Stock Prediction and Trading\thanks{
    Dat Mai, PhD, CFA (\href{mailto:maiydat@gmail.com}{maiydat@gmail.com}) is a quantitative researcher at MKT MediaStats, LLC. I have no conflicts of interest to disclose. The views expressed herein are solely my own and do not reflect those of my employer.  This paper was written as part of my postdoctoral research at the University of Missouri-Columbia. I would like to thank Andrej Karpathy for publicly sharing his lecture and code on the GPT architecture. 
    I acknowledge helpful comments from the participants at the Citi's Data Science Seminar and the 2024 Chicago Quantitative Alliance (CQA) Fall Conference. 
}}
\author{
    Dat Mai
}
\date{
    September 2024 \\
    Click \href{https://papers.ssrn.com/sol3/papers.cfm?abstract_id=4787199}{here} for the most updated version
}

\begin{document}





\maketitle
\thispagestyle{empty}

\begin{abstract}
    \noindent This paper introduces StockGPT, an autoregressive ``number'' model trained and tested on 70 million daily U.S.\ stock returns over nearly 100 years. Treating each return series as a sequence of tokens, StockGPT automatically learns the hidden patterns predictive of future returns via its attention mechanism. On a held-out test sample from 2001 to 2023, daily and monthly rebalanced long-short portfolios formed from StockGPT predictions yield strong performance. The StockGPT-based portfolios span momentum and long-/short-term reversals, eliminating the need for manually crafted price-based strategies, and yield highly significant alphas against leading stock market factors, suggesting a novel AI pricing effect. This highlights the immense promise of generative AI in surpassing human in making complex financial investment decisions.  
    \vspace{10pt}
    
\end{abstract}

\clearpage



\setcounter{page}{1}

\section{Introduction} \label{sec:intro}

Generative artificial intelligence (GenAI)---a set of advanced technologies capable of generating texts, images, videos, programming codes, or arts from instructions via sounds or texts---has taken the society by storm and exerted wide-range influences on many aspects of the world economy (\citealt{noy2023experimental,dell2023navigating,baldassarre2023social,mannuru2023artificial,saetra2023generative,otis2023uneven}). Although it had been around for years, GenAI came to public prominence since the introduction of ChatGPT in November 2022, a chatbox able to generate answers, reasoning, and conversations at human level. 

Since its introduction, ChatGPT and similar large language models have quickly made their ways into the investment industry. One common use of ChatGPT for investment is to give trading recommendations directly from news about a company (such as news articles or corporate communications) (\citealt{lopez2023can}). A less direct approach is to rely on similar pretrained language models such as BERT (\citealt{devlin2018bert}) and OPT (\citealt{zhang2022opt}) to generate a sentiment score for each company which is then used to make trading decisions. For example, \citet{jiang2022expected} and \citet{kirtac2024sentiment} find that stock portfolios formed on sentiment scores generated by BERT and OPT have impressive performance. 

This paper contributes to this fast-evolving field by applying the GenAI logic to numeric stock data. That is, I first train a new Generative Pretrained Transformer (GPT) model (\citealt{brown2020language}) from scratch on numeric stock data (hereafter StockGPT) and then show that StockGPT has the potential to produce strong investment performance.\footnote{
    ChatGPT is GPT finetuned for the conversational purpose.
} 
Unlike previous finance domain-specific language models that are pretrained on financial \textit{texts} such as FinBERT (\citealt{yang2020finbert}) and BloombergGPT (\citealt{wu2023bloomberggpt}), to the best of my knowledge, StockGPT is the first of its kind to be pretrained directly on \textit{numeric} stock return data.

For the trading purpose, using a model trained directly on stock data has three important advantages over models trained on texts: (i) the model learns price patterns directly from price data rather than from news about prices, and (ii) the model predictions are available for each stock at each time point rather than dependent on the availability of news data about stocks, and (iii) the model predicts the whole distribution of future returns rather than a point estimate.

Language models such as GPT operates by predicting the next most likely  token given the previous ones, $p(x_{t+1}|x_t, \dots x_1)$. This nature bears a strong resemblance to numeric time series data such as stock returns where data points come in order and the next value is conditional on what comes before it. Hence the natural question is whether the architecture of language models can be applied to numeric time series data. To do so,  one fundamental difference between texts and numbers needs to be addressed: texts are a collection of (vast but) discrete tokens while numeric time series are generally continuous. Therefore, to train a generative model for stock returns, I first discretize stock return data into intervals (or ``tokens'') and then apply the language model architecture. 

To build the StockGPT model, I adapt a light-weight version of the GPT architecture, which consists of four attention blocks having about one million parameters. Input into the model is a sequence of 256 consecutive daily returns on each stock (i.e., the block size in language models) which approximates the number of trading days in a year.\footnote{
    It is a convention in machine learning to specify model parameters in powers of 2.
} 
The training objective is to predict the next return value given its previous returns using the transformer architecture, which receives indexes (or positions) of the tokens in a sequence, retrieves their vector representations, and models their dependencies via a mechanism called ``attention'' (\citealt{vaswani2017attention}).  The training sample consists of around 50 million daily U.S.\ stock returns from 1926 to 2000, which covers almost all stocks that have ever been listed on the U.S.\ stock market during the 20\textsuperscript{th} century. The model is tested on a hold-out sample of around 20 million daily U.S.\ stock returns from 2001 to 2023. 

Notably, the model is trained only once using the training sample and applied off-the-shelf to the out-of-sample period. This study design serves two purposes: (i) it is the cleanest setup to test the effectiveness of the model and (ii) it reduces the computational costs. Despite this simple setup, the model still delivers strong performance up to 23 years after the period it is trained on. In practice, the model should be continually retrained with the arrival of new financial data to uphold its relevance and performance. This is especially needed in a dynamic environment like the stock market featured by a low signal to noise ratio and constantly distributional shifts (\citealt{kelly2023financial}).

During the testing phase, for each stock on each trading day $t$, StockGPT uses 256 daily returns from $t-255$ to $t$ to make a return forecast for $t+1$. The evaluation of the forecasts consists of two steps. First, I examine the accuracy of the forecasts by running cross sectional regressions of realized stock returns on day $t+1$ onto return predictions for $t+1$. The results indicate that StockGPT makes fairly accurate predictions. 

The second evaluation step entails building real time trading portfolios based on StockGPT forecasts. At the market close of each trading day $t$, I build zero cost portfolios by going long/short the top/bottom decile of stocks having the highest/lowest return forecasts for day $t+1$ and rebalance the portfolio at the $t+1$ market close. To avoid trading only micro stocks since these stocks are illiquid and incur high transaction and market impact costs, before forming the portfolio, I  remove stocks below the 10\textsuperscript{th} percentile market value at the market close.

Under the equal-weighting scheme where each stock receives an equal weight in the portfolio, this daily rebalanced long-short portfolio earns an average annualized returns of 119\% from 2001 to 2023, achieving a Sharpe ratio of 6.5. This performance is higher than the best daily-rebalanced portfolio based on language model predictions  in \citet{jiang2022expected}, which has an annual return of 50\% and Sharpe ratio of 4.8 from 2004 to 2019. It is noteworthy that while the prediction model in \citet{jiang2022expected} is retrained every year,\footnote{
    Specifically, they retrieve contextual word embeddings from pretrained OPT and BERT and use these embeddings to retrain the return prediction model every year. 
} 
StockGPT is  trained only once using data up to 2000. 

Under the value weighting scheme where stock weights in the portfolio are proportional to their market values, the StockGPT-based portfolio achieves an average annualized return of 27\% and a Sharpe ratio of 1. Since value weighting gives more weight to stocks having higher market values, this result is consistent with the consensus view in asset pricing that small stocks are more predictable due to more mispricing and higher arbitrage costs (\citealt{baker2006investor}).


Since StockGPT makes its return forecasts using only historical price data, I examine how it relates to common price-based strategies such as momentum and long-/short-term reversals. I find that the StockGPT-based portfolios span these strategies via the spanning test. This suggests that AI is more effective than human in designing trading strategies based on historical price movements. The StockGPT-based portfolios also encompass several factors of the \citet{fama2015five} five factor model and the \citet{hou2021augmented} q-factor model. 

It is noteworthy that unlike several fields such as medical or law where GenAI is expected to generate (100\%) correct responses, StockGPT does not need to accurately predict future returns on individual stocks for it to become useful. Instead, in the context of cross-sectional asset pricing, it is only required to identify the groups of stocks that are more likely to go up/down to facilitate the long/short trading strategy.


While the daily results show the proof of concept that the GPT model can be applied to numeric stock data to yield strong investment results, it is practically challenging to trade hundreds of stocks on a daily basis, especially the small cap ones. Therefore, I also experiment with a more realistic StockGPT model that makes monthly return predictions. Specifically, I train a new model to predict the returns over the next 20 days instead of the next-day return as before.  I then use this model to make monthly return forecasts and form monthly rebalanced portfolios. 

On average, this strategy earns an annual return of 13\% with a Sharpe ratio of 1 from 2001 to 2023, outperforming 11 common stock factors by a large margin (these factors include momentum, long-/short-term reversals, five factors from \citet{fama2015five}, and three factors from \citet{hou2021augmented}). The performance persists if I focus on only the 50\% largest companies based on market cap or retain only stocks listed on NYSE. The StockGPT portfolio also earns a highly significantly annual alpha of 16\% ($t$-statistic of 4.7) against all of these factors combined, suggesting a new AI-based pricing effect not captured by standard asset pricing factors. In other words, StockGPT can be combined with standard pricing factors to improve the overall risk-return profile.

Although StockGPT shows promising performance, the model can be enhanced in several ways by practitioners to achieve better results. First, the model should be retrained frequently (such as monthly) to maintain its relevance and performance. Second, StockGPT as introduced in this paper is a light-weight adoption of the GPT architecture; it is an open question whether extending the model along several of its parameter dimensions will yield better performance. Third, training StockGPT with high frequency stock data can be a fruitful avenue since there is evidence of alpha to be extracted from the order book (\citealt{kolm2023deep}). These three enhancements can be readily implemented given enough computing power. In addition, one area that needs more exploration and research efforts is how to modify the model itself or the training of it to work better with large cap stocks.

\section{Model Architecture} \label{sec:model}

\subsection{Overview}

\begin{figure}[t]
    \centering
    \includegraphics[width = 0.6\linewidth]{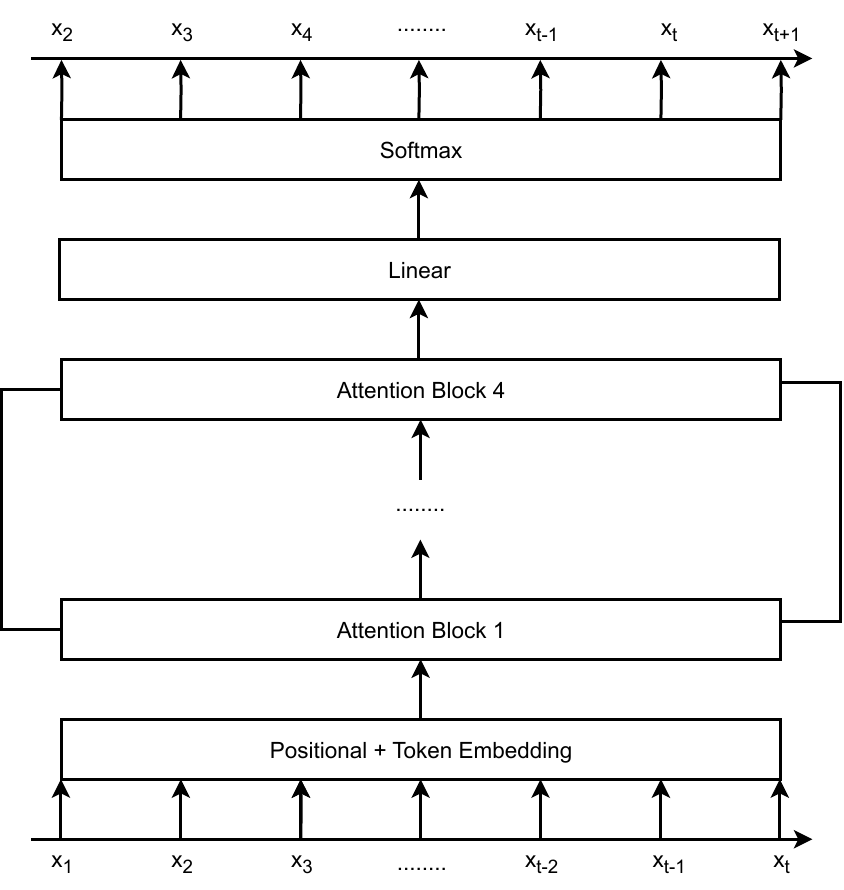}
    \caption{StockGPT Architecture}
    \label{fig:decoder}
\end{figure}

StockGPT uses a vanilla decoder-only transformer architecture which is the second step of the canonical transformer model developed by \citet{vaswani2017attention}. The decoder-only transformer is also the architecture of ChatGPT. \autoref{fig:decoder} depicts the sketch of the architecture. Specifically, the decoder receives an input sequence of tokens $x=(x_1, x_2, \dots, x_{t-1}, x_t)$, transforms its via multiple layers of attention, and outputs the probability of each next token, $p(x_2|x_1), p(x_3|x_2, x_1)$,$\dots$, $p(x_{t+1}|x_t, \dots x_1)$. 

During the training phase, the model learns and updates its parameters via mimimizing the cross-entropy loss  between a token prediction and its actual value $l(\hat{x}_{t+1}, x_{t+1})$ averaging across all tokens across all sequences in a training batch. During the deployment phase, the decoder generates an output sequence of tokens $(x_{t+1}, x_{t+2}, \dots, x_{t+m})$ one at a time, given the input sequence $x$. Specifically, it receives an input sequence $(x_1, x_2, \dots, x_{t-1}, x_t)$, converts it into a conditional probability distribution $p(x_{t+1}|x_t \dots x_1)$, and generates the next token from this distribution. The decoder model is ``autoregressive'' in the sense that it consumes its own generated output at each time as additional input to generate the next one, i.e., $p(x_{t+2}|\hat{x}_{t+1}, x_t, \dots, x_1)$ where $\hat{x}_{t+1}$ is previously generated. 

\subsection{Details}

Since computer itself does not understand human texts, the transformer first quantifies text tokens via  token and positional embedding. \textit{Token} embedding simply retrieves a unique vector representation for each token in a dictionary containing all available tokens. \textit{Positional} embedding vectorizes each token \textit{position} in an input sequence. Without positional embedding, the transformer cannot understand the context and order of tokens. The transformer then sums up token and positional embedding vectors for each token. These embeddings are learnable parameters.
\begin{gather}
\text{embedding} = \text{token embedding} + \text{positional embedding}
\end{gather}
\noindent For example, in the sentence ``The [firm] made a [firm] decision about its capital structure.'' the two [firm] words have the same token embeddings but different positional embeddings due to their positions.

At the heart of the transformer model is the attention mechanism. Accordingly, for each token, the transformer generates three vectors from its embedding vector: key $k$, query $q$, and value $v$. The attention for each token $t$ is the weighted sum of its $v_t$ with all $v_i$'s of tokens preceding it, weighted by the product of its $q_t$ with $k_i$'s of those tokens and a normalizing constant. 
\begin{gather}
\text{attention}_t = \sum_{i=1 \dots t} v_i \times w_i \quad \text{with} \quad w_i = q_t \times k_i \times \text{norm. const.}
\end{gather}
Intuitively, a token emits a query and the previous tokens that can match its query (i.e., having a high $q_t \times k_i$ value) get its attention. $k$, $q$, and $v$ are also learnable parameters.\footnote{
    Technically speaking, the model learns the weight matrices that produce these vectors.
} 
This mechanism constitutes a ``self-attention'' head and helps the transformer develop a contextual understanding of the tokens.\footnote{
    This tutorial by Andrej Karpathy gives an intuitive step-by-step introduction to the GPT architecture: \href{https://www.youtube.com/watch?v=kCc8FmEb1nY\&t=4901s}{https://www.youtube.com/watch?v=kCc8FmEb1nY\&t=4901s}.
} 
In the above example, the attention mechanism helps the model understand that [its] refers to the first [firm]. Since each token is only influenced by the tokens before it, this setup is autoregressive.

The transformer concatenates multiple attention heads into a multi-head node which is sequentially followed by multiple linear layers to form an attention block.\footnote{
    Specifically, for StockGPT, in the first step of the attention block, the input goes through a layer normalization, then multi-head concatenation,  followed by a linear layer with dropout. The output of this layer is added up to the input via a skip connection. In the second step, the output of the first step goes through a second layer normalization, followed by a linear expanding layer to increase the input dimension by 4 times, a ReLU activation, and a contracting layer to revert to the input dimension with dropout. Again the output of this second step is added up to its input via a second skip connection. 
} 
Multiple attention blocks are then stacked on top of each other. The last attention block is follows by a  layer normalization and a linear layer whose output is converted into a vector of probabilities via a softmax activation. Specifically, at time step $t$, the transformer outputs $p(x_{t+1}|x_t \dots x_1)$ which is the multinomial distribution over all available tokens in the dictionary, conditional on all tokens up to $t$. Given this distribution, the model can sample the most likely token at $t+1$ from its dictionary.

\subsection{StockGPT Specifics}

Since the transformer can only work with text tokens, to use it on continuous stock return data, the first step is to discretize returns into intervals. \autoref{tab:return_bin} illustrates the discretization rule. 

Accordingly, I first convert returns into integer basis points by multiplying them by 10,000 and keeping the integer portion. Next, I cut the basis points into intervals of 50, closed on right. The first interval closes on -10,000. Since stock prices cannot be negative, returns cannot be lower than -10,000 basis points (i.e., -100\%); therefore, the first bin contains only -10,000. The last closed interval is (9,950, 10,000]. Third, for each bin, I use the mid value of each interval to represent its value with the exception that the first bin (-Inf, -10,000] is represented by -10,000 and the last bin (10\_000, Inf) by 10,000. In other words, I treat all daily returns greater than 100\% as 100\%. Values above this threshold are extremely rare since the 1\textsuperscript{th} to 99\textsuperscript{th} percentile of daily returns in the training set is from -9.6\% to 11.1\%. Finally, the bins are numbered from 0 to 401. Therefore, my return ``dictionary'' has a total of 402 tokens where each token is a return bin midpoint.  As an example of the discretization rule, the following return sequence (-2.4\%, 0\%, 0\%, 5\%, 4.8\%) is converted into the index sequence (196, 200, 200, 210, 210) which is input into StockGPT.

Besides the vocabulary size of 402, StockGPT has a block size (i.e., length of each input sequence) of 256, token and positional embedding sizes of 128, 4 attention blocks each consisting of 4 self-attention heads, and a dropout probability of 0.2 in various layers. Taken together, StockGPT has 0.93 million parameters. StockGPT is trained in 10,000 training steps with each step consisting of 64 sequences (i.e, batch size) drawn randomly from the training data.\footnote{
    During the training phase, the cross-entropy loss stabilizes at around 2.5 after 5,000 training steps. With 402 labels (the number of return bins), the maximum cross-entropy would be $E = -\sum_{i}log(1/402)\times(1/402)=6$. The model is fully trained locally on a MacBook M2 with 64GB RAM and 30 GPU cores.
}
The probability of sampling each stock during training is proportional to the number of daily return observations it has.

As discussed above, to make a return forecast, given a 256-return sequence input $(x_{t-255}, \dots, x_t)$, StockGPT will output $p(x_{t+1}|x_t, \dots, x_{t-255})$, a multinomial distribution over 402 return bins.\footnote{
    Technically, input of sequence of any length from 1 to 256 (i.e., the block size during training) can be used to make forecasts. Analogously, ChatGPT prompts can be of any length up to a limit (around 2048 tokens). However, since StockGPT is trained with block size of 256, I also use input sequence of 256 days in making forecasts to utilize all price patterns the model has discovered during training. 
}
The model produces output in terms of bin indexes which are converted to numeric returns using the bin midpoints in \autoref{tab:return_bin}. The expected return for day $t+1$ will then be the weighted average of return bin midpoints weighted by the corresponding bin probabilities presented by $p(x_{t+1}|x_t, \dots, x_{t-255})$. Alternatively, the expected returns on day $t+1$ can be computed by sampling many forecasts from $p(x_{t+1}|x_t, \dots, x_{t-255})$ and averaging them. The two approaches will produce the same results if the number of drawn samples is large but the latter approach is more computationally intensive. To make return forecasts over the next $m$ days, we can recursively sample several paths of forecasts $x^j = (x_{t+1}, x_{t+2}, \dots, x_{t+m})$ and average across the paths.

\section{Data} \label{sec:data}

Stock return data comes from Center for Research in Security Prices (CRSP) that collects all historical U.S. stock returns from 1926 to 2023. As standard in asset pricing research, I include only common stocks with a share code of 10 or 11  traded on three main exchanges NYSE, AMEX and NASDAQ. This sample consists of around 70 million stock observations from 1926 to 2023. This sample is then split into two parts: the sample from 1926 to 2000 for training and the sample from 2001 to 2023 for testing. Within the training sample, data from 1926 to 1990 is used for parameter optimization and data from 1991 to 2000 for hyperparameter tuning and evaluation. 

During training evaluation, I document that the model using stocks from NYSE alone has lower evaluation loss than the one using all three exchanges, 2.55 versus 2.72. This may be because NYSE is the world largest stock exchange that lists high quality large cap stocks while AMEX and NASDAQ list smaller stocks that add noises to the training process. Therefore, the main results focus on the model trained on NYSE data alone while the results using the model trained on all three exchanges are reported in \autoref{tab:daily_port_all}.\footnote{
    The latter model still produces annual returns of 83\% with Sharpe ratio of 5.
}

During the testing phase, stock returns from all three exchanges are used. As noted in the introduction, the model is trained only once using the training sample and kept unchanged during the testing phase.

\section{Results: Daily Prediction} \label{sec:results}

\subsection{Fama–MacBeth Regression}

To evaluate the quality of return forecast for day $t+1$, I first compare it against the actual return on that day. Specifically, for each trading day $t$, I run the following cross-sectional regression
\begin{gather}
    x_{it+1} = a_t + b_t \times \hat{x}_{it+1} + e_{it+1}
\end{gather}
where $x_{it+1}$ is the actual realized return of stock $i$ on day $t+1$ and $\hat{x}_{it+1}$ is its StockGPT return forecast. The slope $b_t$ and regression adjusted $R^2_t$ are then averaged across all trading days in the test sample. These measure how well StockGPT forecasts track the actual returns. This regression specification is referred to as the Fama-MacBeth regression in the asset pricing literature (\citealt{fama1973risk}).

\autoref{tab:daily_fmb} reports the results. Accordingly, the average slope coefficient is 0.5, indicating that a cross-sectional difference of 100 basis points (i.e, 1\%) in StockGPT return predictions signals a difference of 50 basis points in realized returns. Moreover, the average cross-sectional $R^2$ is 1.2\% equivalent to a 11\% cross-sectional correlation between return predictions and actual returns. For comparison, the average correlation between return forecasts based on language models and actual returns is around 2\% in \citet{jiang2022expected}. I also examine the relation between return forecasts for day $t+1$ and realized returns on day $t+2$ (i.e., skipping one day). For this test, the slope coefficient is 0.09 and $R^2$ is 0.4\% which translates into a 6\% correlation. The slopes in both tests are highly significant with $t$-statistics over 10. Overall, GPT forecasts track future returns well even after one day.

\subsection{Portfolio Sorting}

\begin{figure}[t]
    \centering
    Panel A: Next Day
    
    \subfloat[\centering High-Low Portfolio]{
        \includegraphics[width = 0.5\linewidth]{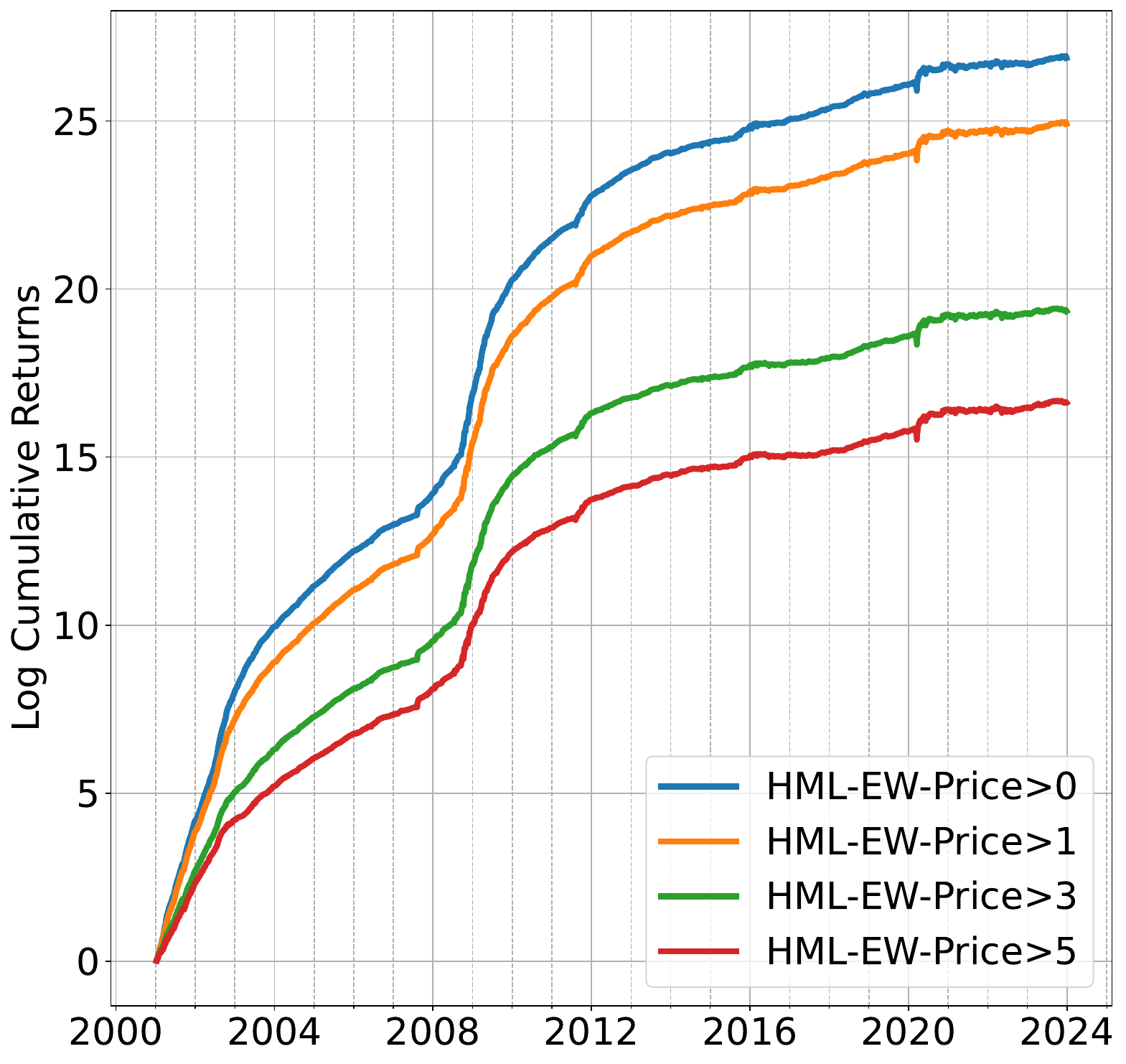}
    }
    \subfloat[\centering High and Low Portfolios]{
        \includegraphics[width = 0.5\linewidth]{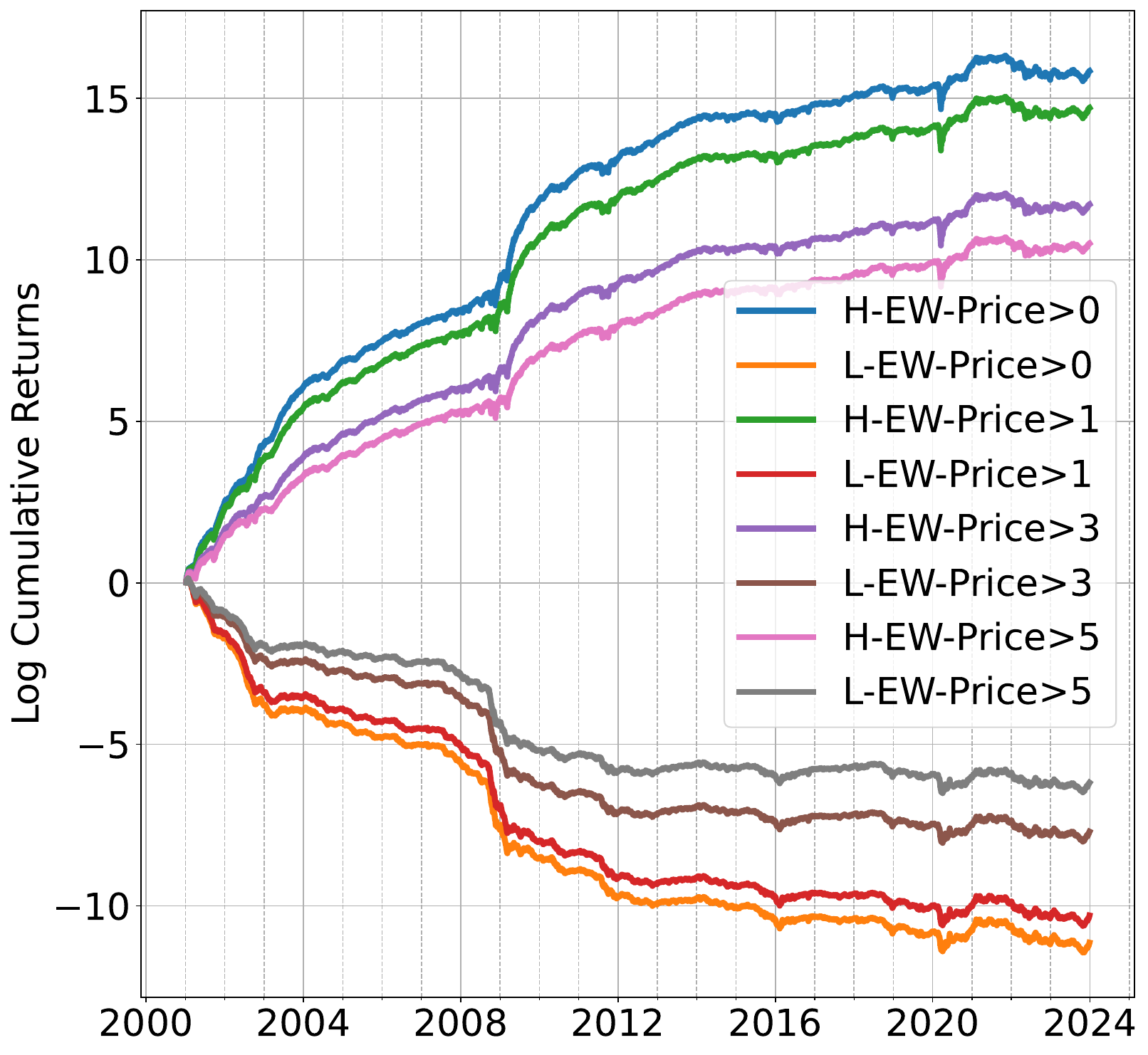}
    } \\

    \vspace{15pt}
    Panel B: Skipping 1 Day
    
    \subfloat[\centering High-Low Portfolio]{
        \includegraphics[width = 0.5\linewidth]{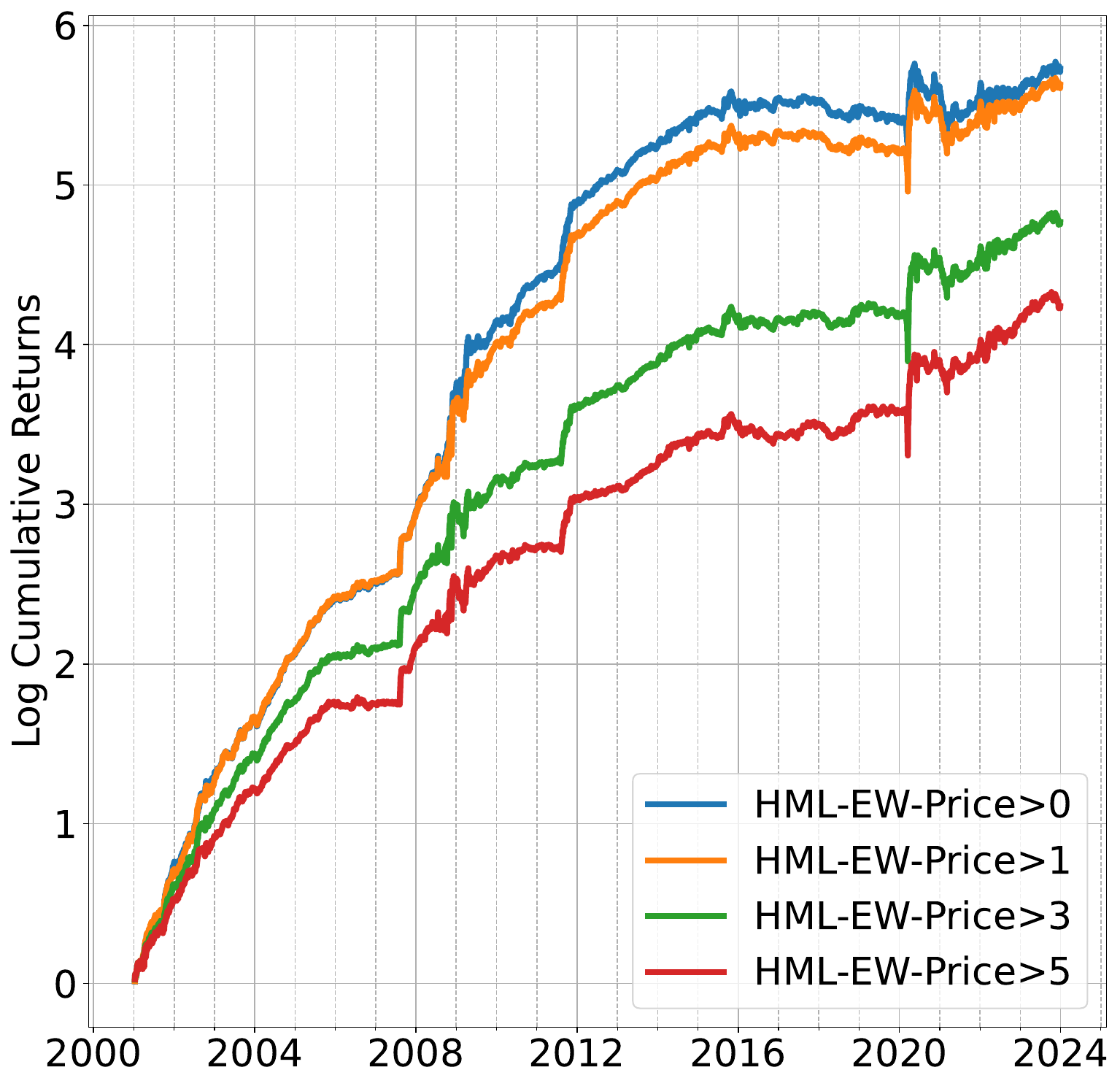}
    }
    \subfloat[\centering High and Low Portfolios]{
        \includegraphics[width = 0.5\linewidth]{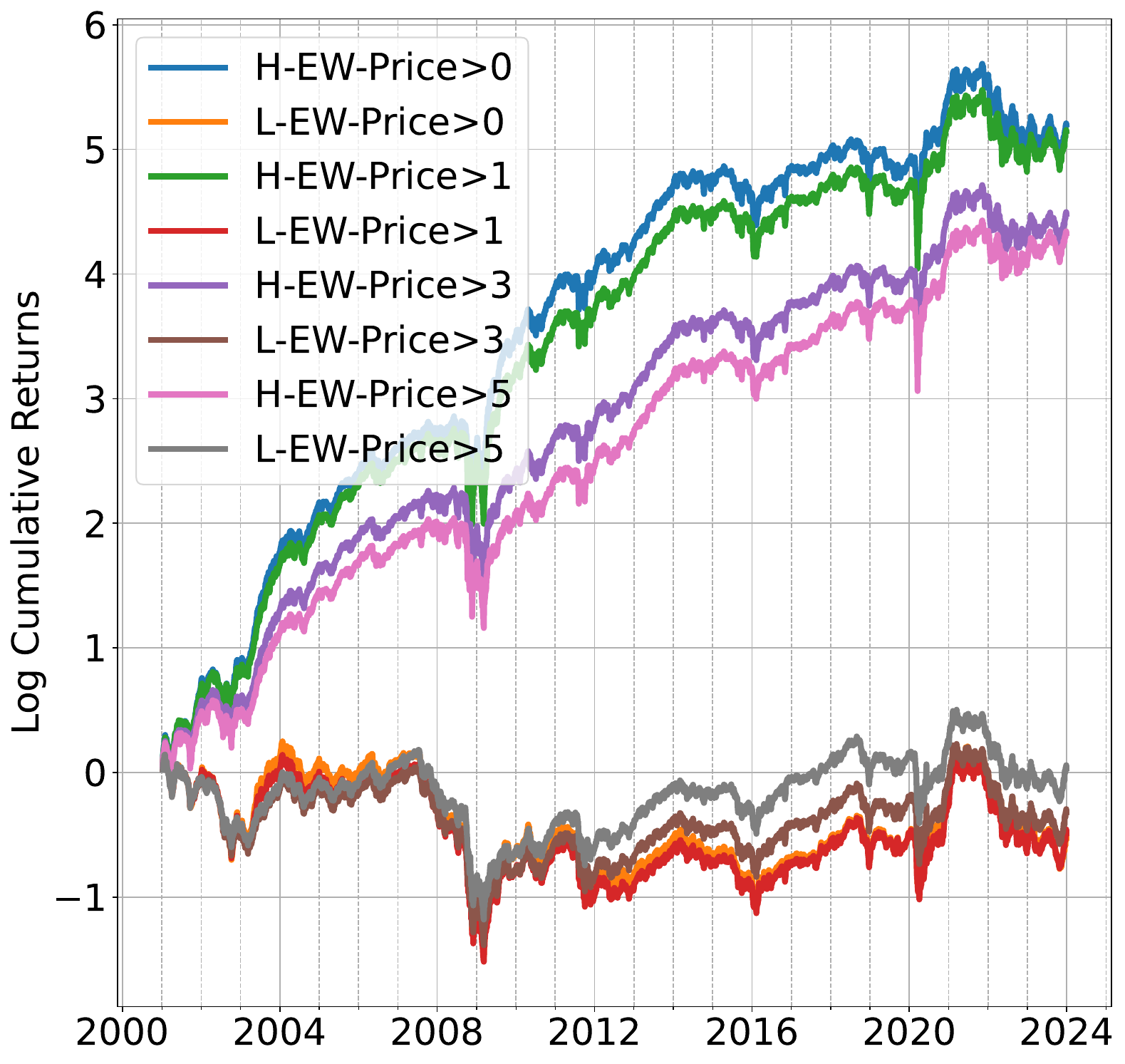}
    } \\
    
    \caption{Daily Cumulative Returns}
    
    \vspace{5pt}
    \justifying
    \footnotesize 
    \noindent This figure plots the \textit{log} cumulative returns of long-short high-minus-low (HML), high (H), and low (L) portfolios formed from StockGPT return forecasts. Portfolios are equal-weighted. Portfolios are formed after excluding stocks in the  bottom decile based on market value. Different portfolios correspond to different market price thresholds under which stocks are excluded. Panel A (B) shows results when return forecasts for day $t+1$ are used to form portfolios for day $t+1$ ($t+2$). The left (right) panel shows results for the long-short (each leg) portfolios. The sample is daily from January 2001 to December 2023.
    
    \label{fig:daily_cumu}
\end{figure}

The main empirical analysis is to examine the trading implications of StockGPT forecasts. To do so, on each trading day $t$, I buy the top stock decile  having the highest return forecasts for $t+1$ (High portfolio) and sell the bottom decile having the lowest return forecasts for $t+1$ (Low portfolio). To avoid only trading micro-cap stocks, I remove stocks having market value below the 10\textsuperscript{th} percentile each day. In \autoref{tab:daily_port}, under equal weighting (EW), this long-short portfolio yields an average annual return of 119\% with a Sharpe ratio (mean/standard deviation) of 6.5.\footnote{
    Without the market cap restriction, the StockGPT-based portfolio would earns 230\% annually with a Sharpe ratio of 10. On the other hand, if stocks having market cap below the 30\textsuperscript{th} percentile are removed, the resulting portfolio earns 50\% annually with a Sharpe ratio of 2.9.
}  
If I remove stocks having prices below \$1, \$3, and \$5, the mean returns (and Sharpe ratios) are 110\% (6.3), 86\% (5.2), and 74\% (4.7), respectively. The left graph in Panel A of \autoref{fig:daily_cumu} plots the \textit{log} cumulative returns of these 4 long-short portfolios. These portfolios show a consistent upward trend throughout the 2001-2023 sample with the biggest jump in 2009 after the financial crisis. The right graph in Panel A plots the cumulative returns of each long/short leg of the portfolios. It is clear that StockGPT can symmetrically predict both rising and falling stocks.

While the annual return of 119\% under equal weighting in the baseline model is before transaction costs, under the hypothetical worst-case scenario that the portfolio replaces all of its constituents every day (i.e., a turnover of 400\% in a long-short portfolio) and each trade costs 5 basis points, the StockGPT-based strategy still realizes an annual return of  69\% net of transaction costs.

Under value weighting (VW), the long-short portfolio without price filter (but still after removing the bottom decile based on market cap) earns an annual returns of 27\% and Sharpe ratio of 1. The Sharpe ratio of the portfolios with price filters are at 1, 0.9, and 0.8 for the \$1, \$3, and \$5 price thresholds, respectively. Since value weighting gives more weight to large cap stocks, this result indicates that StockGPT is more effective at forecasting returns of small cap stocks. This is expected since small cap stocks are more likely to be mispriced (\citealt{baker2006investor}).

\autoref{tab:daily_port} also reports the portfolio when return forecasts for $t+1$ are used to form portfolio for $t+2$. Under equal weighting, this skipping-one-day portfolio earns 26\% annually with a Sharpe ratio of 1.7. Panel B of \autoref{fig:daily_cumu} shows that when one day is skipped, StockGPT forecasts track the returns in the long leg better than the short one. 

\subsection{Spanning Test}

Since StockGPT uses only historical market price data to make return forecasts, it is important to examine how the StockGPT-based portfolio relates to prominent trading strategies based on historical returns. The three most notable patterns are short-term reversal (using returns from month $t-1$) by \citet{jegadeesh1990evidence}, momentum (using returns from month $t-2$ to $t-12$) by \citet{jegadeesh1993returns}, and long-term reversal (using returns from month $t-13$ to $t-60$) by \citet{de1985does}. It is also interesting to examine how StockGPT performs relative to leading stock factors such as the five factors of \citet{fama2015five} and the investment-based q5 factors of \citet{hou2021augmented}.\footnote{
    The momentum, reversal, and five factors of \citet{fama2015five} are available at \href{https://mba.tuck.dartmouth.edu/pages/faculty/ken.french/data_library.html}{https://mba.tuck.dartmouth.edu/pages/faculty/ken.french/data\_library.html} while the q5 factors of \citet{hou2021augmented} are at \href{https://global-q.org/factors.html}{https://global-q.org/factors.html}.
} 

As standard in asset pricing research, to examine whether a strategy earns abnormal returns relative to a set of other traded factors, we can perform the following contemporaneous regression:
\begin{gather}
    y_t = \alpha + \beta \times x_t + e_t
\end{gather}
where $y_t$ is the return of the target strategy and $x_t$ is the set of benchmark factors. If $\alpha$ is significant, then $y_t$ earns abnormal returns relative to $x_t$; otherwise, $y_t$ is spanned or encompassed by $x_t$. This test is also referred to as the spanning test.

Panel A of \autoref{tab:alpha} reports the results of the spanning tests in which $y_t$ is the daily returns on the StockGPT-based portfolios and $x_t$ is the set of benchmark factors. Accordingly, both the equal-weighted and value-weighted StockGPT portfolios earn sizable and highly significant alphas relative to all 11 benchmark factors ($t$-statistics are greater than 10 for EW and greater than 3 for VW). 

In Panel B, I test whether the StockGPT portfolios span the benchmark factors. The equal-weighted StockGPT portfolio spans momentum, long-term reversal, value, and size while the value-weighted StockGPT portfolio spans 9 out of 11 factors except profitability from Fama-French and earning growth from q5 models. That the value-weighted StockGPT portfolio better spans the other factors than does the equal-weighted StockGPT portfolios is expected since those factors are also value-weighted.

Overall, the spanning tests show that when we let the stock data speaks for itself via the attention mechanism in StockGPT, handcrafted price-based strategies such as short-term reversal, momentum, and long-term reversal are no longer needed. Notably, although StockGPT only learns from historical returns over the past 12 months, it completely encompasses the long-term reversal pattern based on returns beyond the past 12 months. Furthermore, StockGPT-based portfolios also encompass most leading stock factors.

\section{Results: Monthly Prediction} \label{sec:monthly}

While the daily results show proof of concept that StockGPT can deliver strong investment performance, it is costly and challenging to trade hundreds of stocks on a daily basis. In this section, I examine the arising question of whether StockGPT can be used to make longer term forecasts to build lower frequency portfolios.

There are two ways to produce long-term return forecasts over the next $m$ days from StockGPT. The first approach it to produce several paths of forecasts $x^j = (x_{t+1}, x_{t+2}, \dots, x_{t+m})$ and average across the paths to compute the expected returns over the next $m$ days. However, this approach is very computationally expensive since there are on average 3,000 to 4,000 stocks traded in the cross section. For each stock on each rebalance day, we need to make many $m$-day forecast paths and average them.

The second approach is to train a new StockGPT model where the training target is to predict return over the next $m$ days (i.e., $p(\bar{x}_{t+1 \rightarrow t+m}|x_t, \dots, x_{t-255})$ where $\bar{x}_{t+1 \rightarrow t+m}$ is the mean return over the next $m$ days). I pursue this approach in this section. Specifically, I train a StockGPT model using historical returns to predict mean returns over the next 20 days (i.e., 20 days approximates the number of trading days in a month) with all other specifications kept unchanged from the daily model. Like before, the model is trained only once using data up to 2000. During the testing phase, at the end of each month for each stock, an input sequence of 256 previous daily returns for that stock is used by the new StockGPT model to predict the next 20-day returns (i.e., return over the next month). 

\subsection{Fama–MacBeth Regression}

To evaluate the quality of long-term forecasts by StockGPT, I regress realized monthly returns onto 20-day return forecasts via the Fama-McBeth test discussed above. As reported in \autoref{tab:monthly_fmb}, the average slope coefficient is 3 (significant at 5\%), indicating that a cross-sectional difference of 100 basis points (i.e, 1\%) in StockGPT return forecasts signals a difference of 300 basis points in realized returns. Moreover, the average cross-sectional $R^2$ is 0.55\% equivalent to a 7.4\% cross-sectional correlation between return predictions and actual returns. When one month is skipped between 20-day return forecasts and realized returns, the correlation shrinks to zero.

\subsection{Portfolio Sorting}

I then form monthly rebalanced long-short decile portfolios using the 20-day return forecasts and report the performance statistics over 2001-2023 in Panel A of \autoref{tab:monthly_port}. The equal-weighted portfolios after removing the bottom stock decile based on market value earn about 13\% annually, significant at 1\%, with Sharpe ratios around 1. To ensure the tradability of the strategy, I further remove stocks in the bottom 30\textsuperscript{th} percentile and the performance remains almost unchanged. When the bottom half of all stocks are removed, the annual mean return falls to about 10\% with the Sharpe ratio falling to about 0.7. When only NYSE stocks are used, the average annual returns are 15\% with a Sharpe ratio of 0.9.

For comparison, in Panel B of \autoref{tab:monthly_port}, I report the summary statistics for 11 common stock factors. Only 5 factors yield significant returns, including short-term reversal, market, profitability from \citet{fama2015five}, and return on equity and earnings growth from \citet{hou2021augmented}. Among these factors, short-term reversal yields the strongest result with a mean return of 8.8\% and a Sharpe ratio of 0.7. It is clear that StockGPT portfolios across different specifications outperform these factors. 

\begin{figure}[t]
    \centering
    Panel A: StockGPT Portfolios
    
    \subfloat[\centering High-Low Portfolio]{
        \includegraphics[width = 0.5\linewidth]{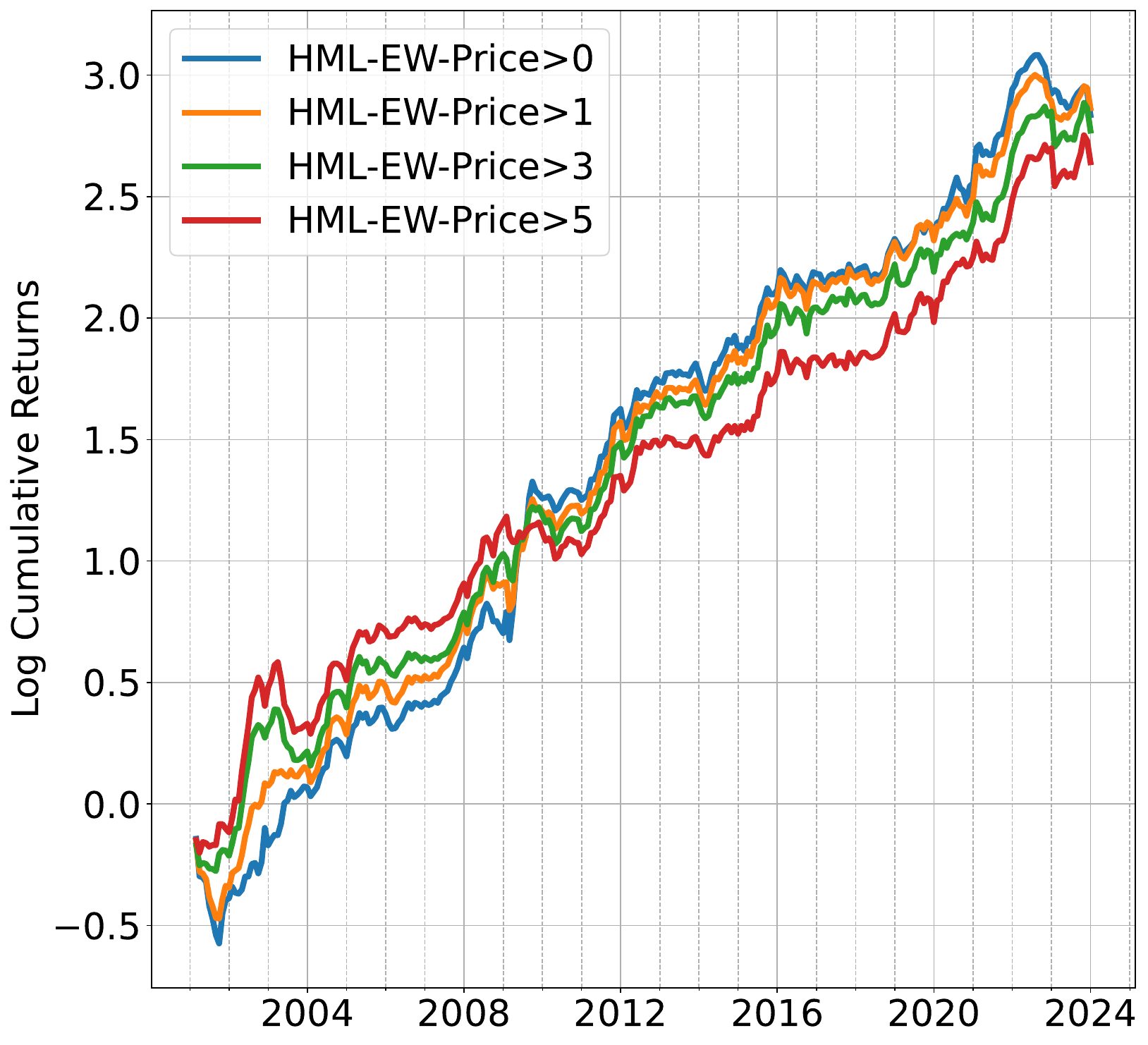}
    }
    \subfloat[\centering High and Low Portfolios]{
        \includegraphics[width = 0.5\linewidth]{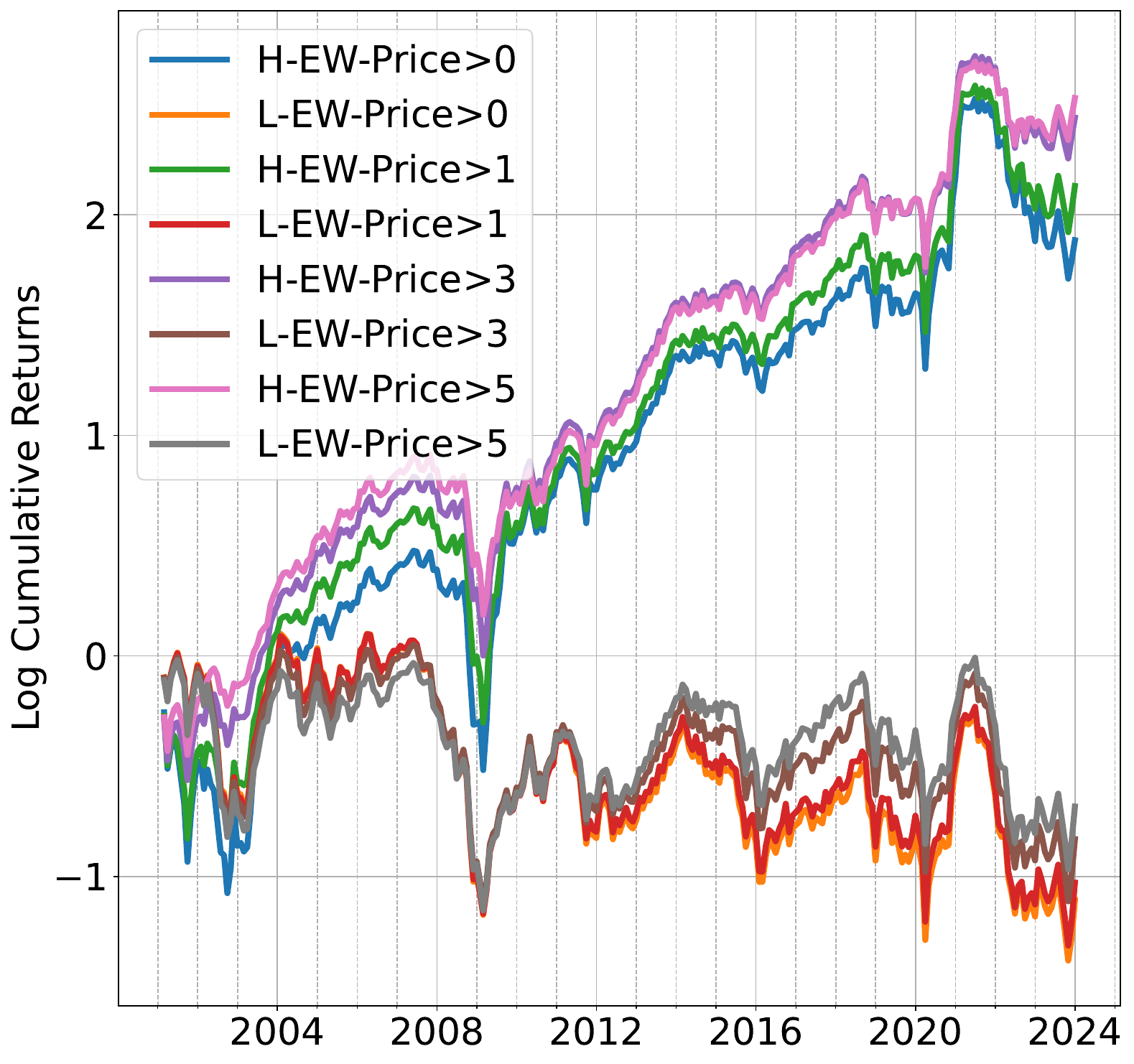}
    } \\

    \vspace{15pt}
    Panel B: Stock Factors
    \includegraphics[width = 0.9\linewidth]{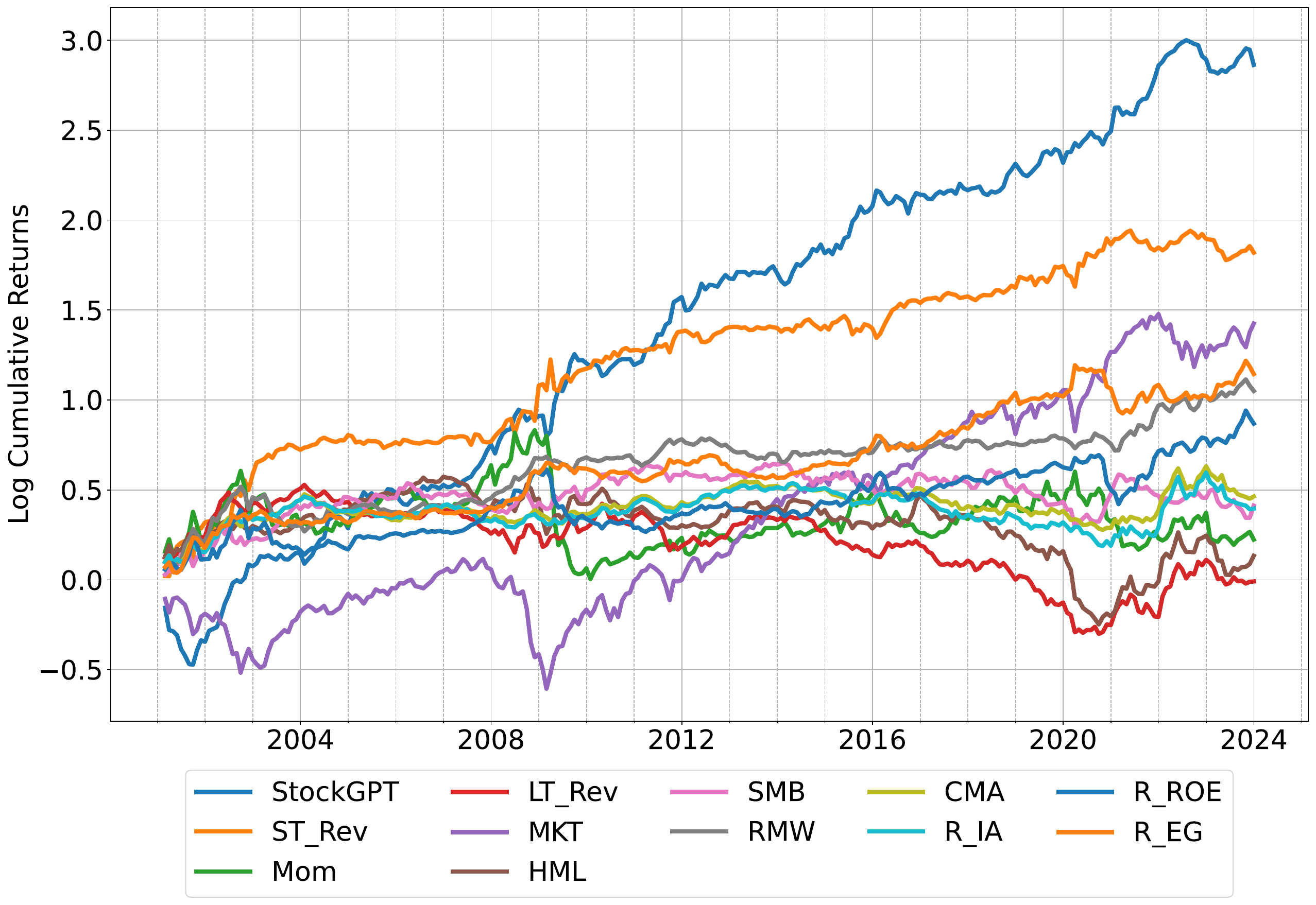}
    
    \caption{Monthly Cumulative Returns}

    \vspace{5pt}
    \justifying
    \footnotesize 
    \noindent Panel A plots the log cumulative returns of long-short high-minus-low (HML), high (H), and low (L) portfolios formed from StockGPT return forecasts. Portfolios are equal-weighted. Portfolios are formed after excluding stocks in the  bottom decile based on market value. Different portfolios correspond to different market price thresholds under which stocks are excluded. Panel B plots the log cumulative returns of StockGPT-based portfolio and stock factors, including short-term reversal (ST\_Rev), momentum (Mom), long-term reversal (LT\_Rev), market (MKT), value (HML), size (SMB), profitability (RMW), and investment (CMA) from \citet{fama2015five}, as well as investment (R\_IA), return on equity (R\_ROE), and earnings growth (R\_EG) from \citet{hou2021augmented}. The sample is monthly from February 2001 to December 2023.
    
    \label{fig:monthly_cumu}
\end{figure}

Panel A of \autoref{fig:monthly_cumu} plots the \textit{log} cumulative returns on the StockGPT portfolios. The long-short portfolios see a stable upward trend from 2001 to 2023. Between the two legs of the strategy, StockGPT does better at predicting the future winners. Panel B plots the log cumulative returns of the baseline StockGPT portfolio  and 11 stock factors. Among the factors, short-term reversals outperformed StockGPT before the financial crisis but has lagged StockGPT by a large extent since then.

\subsection{Spanning Test}

\autoref{tab:monthly_alpha} reports the spanning tests. In Panel A, the monthly-rebalanced equal-weighted StockGPT portfolio earns a significant annual alpha of about 15\% ($t$-statistic of 4.7) against all of the factors. This suggests that StockGPT represents a new AI pricing effect not captured by standard factor models. In Panel B, I check whether the stock factors earn alphas against StockGPT. Accordingly, long-term reversal, value, and investment (from both \citet{fama2015five} and \citet{hou2021augmented}) are subsumed by StockGPT. Besides, the market alpha against StockGPT is only marginally significant at the 10\% level.

Overall, while the investment performance of the monthly StockGPT model is far less impressive than that of the daily model, it still outperforms all of the standard stock factors and yields highly significant alphas. The monthly results confirm that StockGPT can be used in practice to implement tradable strategies.

\section{Conclusion}

This paper introduces StockGPT, a decoder-only transformer model trained directly on U.S.\ stock returns. Instead of relying on manually crafted trading patterns using historical stock prices, StockGPT automatically learns the hidden patterns most predictive of future returns via its attention mechanism. Even though trained on daily returns only up to 2000, StockGPT delivers strong performance up to 23 years later. The StockGPT-based portfolios encompass common price-based trading strategies such as momentum and long-/short-term reversals and span several leading stock factors such as market, size, value, and investment. 

StockGPT can be enhanced in several ways. First, StockGPT should be retrained frequently with the arrival of new stock data to maintain its performance. Second, StockGPT as introduced in this paper is a light-weight model with only around one million parameters. The natural extension is to examine bigger models having more granular return intervals, longer block size, bigger embedding size, and more layers of attention blocks. Third, examining the long-term forecasts from the daily StockGPT model (as discussed in \autoref{sec:monthly}) can be a fruitful direction. Finally, training StockGPT with higher frequency data such as tick size data can yield promising results.

\clearpage


\begingroup
\setstretch{1}
\setlength\bibitemsep{5pt}
\printbibliography[title={References}]
\endgroup

\clearpage


\clearpage

\begin{table}[t]
    \caption{Return Bins}
    
    \vspace{5pt}
    \footnotesize
    \noindent This table shows the return bins in basis points, bin midpoints, and the corresponding bin indexes.
    \vspace{5pt}
    
    \centering
    
  \providecommand{\huxb}[2]{\arrayrulecolor[RGB]{#1}\global\arrayrulewidth=#2pt}
  \providecommand{\huxvb}[2]{\color[RGB]{#1}\vrule width #2pt}
  \providecommand{\huxtpad}[1]{\rule{0pt}{#1}}
  \providecommand{\huxbpad}[1]{\rule[-#1]{0pt}{#1}}
\begin{tabular}{l l l}

\hhline{>{\huxb{0, 0, 0}{0.8}}->{\huxb{0, 0, 0}{0.8}}->{\huxb{0, 0, 0}{0.8}}-}
\arrayrulecolor{black}

\multicolumn{1}{!{\huxvb{0, 0, 0}{0}}c!{\huxvb{0, 0, 0}{0}}}{\huxtpad{0pt + 1em}\centering \hspace{6pt} Return Bin \hspace{6pt}\huxbpad{0pt}} &
\multicolumn{1}{c!{\huxvb{0, 0, 0}{0}}}{\huxtpad{0pt + 1em}\centering \hspace{6pt} Bin Midpoint \hspace{6pt}\huxbpad{0pt}} &
\multicolumn{1}{c!{\huxvb{0, 0, 0}{0}}}{\huxtpad{0pt + 1em}\centering \hspace{6pt} Bin Index \hspace{6pt}\huxbpad{0pt}} \tabularnewline[-0.5pt]

\hhline{>{\huxb{0, 0, 0}{0.8}}->{\huxb{0, 0, 0}{0.8}}->{\huxb{0, 0, 0}{0.8}}-}
\arrayrulecolor{black}

\multicolumn{1}{!{\huxvb{0, 0, 0}{0}}c!{\huxvb{0, 0, 0}{0}}}{\huxtpad{0pt + 1em}\centering \hspace{6pt} (-Inf, -10,000] \hspace{6pt}\huxbpad{0pt}} &
\multicolumn{1}{c!{\huxvb{0, 0, 0}{0}}}{\huxtpad{0pt + 1em}\centering \hspace{6pt} -10\_000 \hspace{6pt}\huxbpad{0pt}} &
\multicolumn{1}{c!{\huxvb{0, 0, 0}{0}}}{\huxtpad{0pt + 1em}\centering \hspace{6pt} 0 \hspace{6pt}\huxbpad{0pt}} \tabularnewline[-0.5pt]

\hhline{}
\arrayrulecolor{black}

\multicolumn{1}{!{\huxvb{0, 0, 0}{0}}c!{\huxvb{0, 0, 0}{0}}}{\huxtpad{0pt + 1em}\centering \hspace{6pt} (-10\_000, -9\_950] \hspace{6pt}\huxbpad{0pt}} &
\multicolumn{1}{c!{\huxvb{0, 0, 0}{0}}}{\huxtpad{0pt + 1em}\centering \hspace{6pt} -9\_975 \hspace{6pt}\huxbpad{0pt}} &
\multicolumn{1}{c!{\huxvb{0, 0, 0}{0}}}{\huxtpad{0pt + 1em}\centering \hspace{6pt} 1 \hspace{6pt}\huxbpad{0pt}} \tabularnewline[-0.5pt]

\hhline{}
\arrayrulecolor{black}

\multicolumn{1}{!{\huxvb{0, 0, 0}{0}}c!{\huxvb{0, 0, 0}{0}}}{\huxtpad{0pt + 1em}\centering \hspace{6pt} (-9\_950, 9\_900] \hspace{6pt}\huxbpad{0pt}} &
\multicolumn{1}{c!{\huxvb{0, 0, 0}{0}}}{\huxtpad{0pt + 1em}\centering \hspace{6pt} -9\_925 \hspace{6pt}\huxbpad{0pt}} &
\multicolumn{1}{c!{\huxvb{0, 0, 0}{0}}}{\huxtpad{0pt + 1em}\centering \hspace{6pt} 2 \hspace{6pt}\huxbpad{0pt}} \tabularnewline[-0.5pt]

\hhline{}
\arrayrulecolor{black}

\multicolumn{1}{!{\huxvb{0, 0, 0}{0}}c!{\huxvb{0, 0, 0}{0}}}{\huxtpad{0pt + 1em}\centering \hspace{6pt} ... \hspace{6pt}\huxbpad{0pt}} &
\multicolumn{1}{c!{\huxvb{0, 0, 0}{0}}}{\huxtpad{0pt + 1em}\centering \hspace{6pt} ... \hspace{6pt}\huxbpad{0pt}} &
\multicolumn{1}{c!{\huxvb{0, 0, 0}{0}}}{\huxtpad{0pt + 1em}\centering \hspace{6pt} ... \hspace{6pt}\huxbpad{0pt}} \tabularnewline[-0.5pt]

\hhline{}
\arrayrulecolor{black}

\multicolumn{1}{!{\huxvb{0, 0, 0}{0}}c!{\huxvb{0, 0, 0}{0}}}{\huxtpad{0pt + 1em}\centering \hspace{6pt} (9\_950, 10\_000] \hspace{6pt}\huxbpad{0pt}} &
\multicolumn{1}{c!{\huxvb{0, 0, 0}{0}}}{\huxtpad{0pt + 1em}\centering \hspace{6pt} 9\_975 \hspace{6pt}\huxbpad{0pt}} &
\multicolumn{1}{c!{\huxvb{0, 0, 0}{0}}}{\huxtpad{0pt + 1em}\centering \hspace{6pt} 400 \hspace{6pt}\huxbpad{0pt}} \tabularnewline[-0.5pt]

\hhline{}
\arrayrulecolor{black}

\multicolumn{1}{!{\huxvb{0, 0, 0}{0}}c!{\huxvb{0, 0, 0}{0}}}{\huxtpad{0pt + 1em}\centering \hspace{6pt} (10\_000, +Inf) \hspace{6pt}\huxbpad{0pt}} &
\multicolumn{1}{c!{\huxvb{0, 0, 0}{0}}}{\huxtpad{0pt + 1em}\centering \hspace{6pt} 10\_000 \hspace{6pt}\huxbpad{0pt}} &
\multicolumn{1}{c!{\huxvb{0, 0, 0}{0}}}{\huxtpad{0pt + 1em}\centering \hspace{6pt} 401 \hspace{6pt}\huxbpad{0pt}} \tabularnewline[-0.5pt]

\hhline{>{\huxb{0, 0, 0}{0.8}}->{\huxb{0, 0, 0}{0.8}}->{\huxb{0, 0, 0}{0.8}}-}
\arrayrulecolor{black}
\end{tabular}
    \label{tab:return_bin}
\end{table}

\begin{table}[t]
    \caption{Daily Fama-MacBeth Regression}

    \vspace{5pt}
    \footnotesize 
    \noindent This table reports the time series averages of slopes and adjusted $R^2$'s of the following cross-sectional regression
    \begin{gather*}
    x_{it+1} = a_t + b_t \times \hat{x}_{it+1} + e_{it+1}
    \end{gather*}
    where $x_{it+1}$ is the actual realized return of stock $i$ on day $t+1$ and $\hat{x}_{it+1}$ is its StockGPT return forecast. Returns are in basis points and $R^2$ in percentage points. $t$ is the $t$-statistic of time-series mean of $b_t$ computed using \citet{newey1987hypothesis} standard error with 20 lags. Horizon 1 (2) means comparing return forecasts for $t+1$ with actual returns on $t+1$ ($t+2$). The sample is daily from January 2001 to December 2023.
    \vspace{5pt}
    
    \centering
    
  \providecommand{\huxb}[2]{\arrayrulecolor[RGB]{#1}\global\arrayrulewidth=#2pt}
  \providecommand{\huxvb}[2]{\color[RGB]{#1}\vrule width #2pt}
  \providecommand{\huxtpad}[1]{\rule{0pt}{#1}}
  \providecommand{\huxbpad}[1]{\rule[-#1]{0pt}{#1}}
\begin{tabular}{l l l l}

\hhline{>{\huxb{0, 0, 0}{0.8}}->{\huxb{0, 0, 0}{0.8}}->{\huxb{0, 0, 0}{0.8}}->{\huxb{0, 0, 0}{0.8}}-}
\arrayrulecolor{black}

\multicolumn{1}{!{\huxvb{0, 0, 0}{0}}c!{\huxvb{0, 0, 0}{0}}}{\huxtpad{0pt + 1em}\centering \hspace{6pt} Horizon \hspace{6pt}\huxbpad{0pt}} &
\multicolumn{1}{c!{\huxvb{0, 0, 0}{0}}}{\huxtpad{0pt + 1em}\centering \hspace{6pt} b \hspace{6pt}\huxbpad{0pt}} &
\multicolumn{1}{c!{\huxvb{0, 0, 0}{0}}}{\huxtpad{0pt + 1em}\centering \hspace{6pt} $t$ \hspace{6pt}\huxbpad{0pt}} &
\multicolumn{1}{c!{\huxvb{0, 0, 0}{0}}}{\huxtpad{0pt + 1em}\centering \hspace{6pt} $R^2$ \hspace{6pt}\huxbpad{0pt}} \tabularnewline[-0.5pt]

\hhline{>{\huxb{0, 0, 0}{0.8}}->{\huxb{0, 0, 0}{0.8}}->{\huxb{0, 0, 0}{0.8}}->{\huxb{0, 0, 0}{0.8}}-}
\arrayrulecolor{black}

\multicolumn{1}{!{\huxvb{0, 0, 0}{0}}c!{\huxvb{0, 0, 0}{0}}}{\huxtpad{0pt + 1em}\centering \hspace{6pt} 1 \hspace{6pt}\huxbpad{0pt}} &
\multicolumn{1}{c!{\huxvb{0, 0, 0}{0}}}{\huxtpad{0pt + 1em}\centering \hspace{6pt} 0.50 \hspace{6pt}\huxbpad{0pt}} &
\multicolumn{1}{c!{\huxvb{0, 0, 0}{0}}}{\huxtpad{0pt + 1em}\centering \hspace{6pt} 25.18 \hspace{6pt}\huxbpad{0pt}} &
\multicolumn{1}{c!{\huxvb{0, 0, 0}{0}}}{\huxtpad{0pt + 1em}\centering \hspace{6pt} 1.19 \hspace{6pt}\huxbpad{0pt}} \tabularnewline[-0.5pt]

\hhline{}
\arrayrulecolor{black}

\multicolumn{1}{!{\huxvb{0, 0, 0}{0}}c!{\huxvb{0, 0, 0}{0}}}{\huxtpad{0pt + 1em}\centering \hspace{6pt} 2 \hspace{6pt}\huxbpad{0pt}} &
\multicolumn{1}{c!{\huxvb{0, 0, 0}{0}}}{\huxtpad{0pt + 1em}\centering \hspace{6pt} 0.09 \hspace{6pt}\huxbpad{0pt}} &
\multicolumn{1}{c!{\huxvb{0, 0, 0}{0}}}{\huxtpad{0pt + 1em}\centering \hspace{6pt} 10.20 \hspace{6pt}\huxbpad{0pt}} &
\multicolumn{1}{c!{\huxvb{0, 0, 0}{0}}}{\huxtpad{0pt + 1em}\centering \hspace{6pt} 0.41 \hspace{6pt}\huxbpad{0pt}} \tabularnewline[-0.5pt]

\hhline{>{\huxb{0, 0, 0}{0.8}}->{\huxb{0, 0, 0}{0.8}}->{\huxb{0, 0, 0}{0.8}}->{\huxb{0, 0, 0}{0.8}}-}
\arrayrulecolor{black}
\end{tabular}
    \label{tab:daily_fmb}
\end{table}

\begin{table}[t]
    \caption{Daily Portfolio Statistics}

    \vspace{5pt}
    \footnotesize
    \noindent This table reports the return statistics of the daily long-short StockGPT-based portfolios. Mean and SD (standard deviation) are in annualized percentage points; Mean/SD (Sharpe ratio) is annualized; Min, Max, and MDD (max drawdown) are in percentage points; and t-Mean is $t$-statistic of the mean portfolio return using Newey-West standard error with 20 lags. Portfolios are formed after excluding stocks in the bottom decile based on market value. Horizon 1 (2) refers to using return forecasts for day $t+1$ to form portfolios for day $t+1$ ($t+2$). EW (VW) refers to equal-weighting (value weighting). Price Filter refers to the price level under which stocks are removed. The sample is daily from January 2001 to December 2023.
    \vspace{5pt}
    
    \centering
    \begin{adjustbox}{width = \columnwidth, center}
        
  \providecommand{\huxb}[2]{\arrayrulecolor[RGB]{#1}\global\arrayrulewidth=#2pt}
  \providecommand{\huxvb}[2]{\color[RGB]{#1}\vrule width #2pt}
  \providecommand{\huxtpad}[1]{\rule{0pt}{#1}}
  \providecommand{\huxbpad}[1]{\rule[-#1]{0pt}{#1}}
\begin{tabular}{l l l l l l l l l l}

\hhline{>{\huxb{0, 0, 0}{0.8}}->{\huxb{0, 0, 0}{0.8}}->{\huxb{0, 0, 0}{0.8}}->{\huxb{0, 0, 0}{0.8}}->{\huxb{0, 0, 0}{0.8}}->{\huxb{0, 0, 0}{0.8}}->{\huxb{0, 0, 0}{0.8}}->{\huxb{0, 0, 0}{0.8}}->{\huxb{0, 0, 0}{0.8}}->{\huxb{0, 0, 0}{0.8}}-}
\arrayrulecolor{black}

\multicolumn{1}{!{\huxvb{0, 0, 0}{0}}c!{\huxvb{0, 0, 0}{0}}}{\huxtpad{0pt + 1em}\centering \hspace{0pt} Horizon \hspace{0pt}\huxbpad{0pt}} &
\multicolumn{1}{c!{\huxvb{0, 0, 0}{0}}}{\huxtpad{0pt + 1em}\centering \hspace{0pt} Weight \hspace{0pt}\huxbpad{0pt}} &
\multicolumn{1}{c!{\huxvb{0, 0, 0}{0}}}{\huxtpad{0pt + 1em}\centering \hspace{0pt} Price Filter \hspace{0pt}\huxbpad{0pt}} &
\multicolumn{1}{c!{\huxvb{0, 0, 0}{0}}}{\huxtpad{0pt + 1em}\centering \hspace{0pt} Mean \hspace{0pt}\huxbpad{0pt}} &
\multicolumn{1}{c!{\huxvb{0, 0, 0}{0}}}{\huxtpad{0pt + 1em}\centering \hspace{0pt} t-Mean \hspace{0pt}\huxbpad{0pt}} &
\multicolumn{1}{c!{\huxvb{0, 0, 0}{0}}}{\huxtpad{0pt + 1em}\centering \hspace{0pt} SD \hspace{0pt}\huxbpad{0pt}} &
\multicolumn{1}{c!{\huxvb{0, 0, 0}{0}}}{\huxtpad{0pt + 1em}\centering \hspace{0pt} Mean/SD \hspace{0pt}\huxbpad{0pt}} &
\multicolumn{1}{c!{\huxvb{0, 0, 0}{0}}}{\huxtpad{0pt + 1em}\centering \hspace{0pt} Min \hspace{0pt}\huxbpad{0pt}} &
\multicolumn{1}{c!{\huxvb{0, 0, 0}{0}}}{\huxtpad{0pt + 1em}\centering \hspace{0pt} Max \hspace{0pt}\huxbpad{0pt}} &
\multicolumn{1}{c!{\huxvb{0, 0, 0}{0}}}{\huxtpad{0pt + 1em}\centering \hspace{0pt} MDD \hspace{0pt}\huxbpad{0pt}} \tabularnewline
\end{tabular}
    \end{adjustbox}
    
    \label{tab:daily_port}
\end{table}

\begin{table}[t]
    \caption{Daily Spanning Tests}

    \vspace{5pt}
    \footnotesize
    \noindent This table reports results of the following spanning test
    \begin{gather*}
        y_t = \alpha + \beta \times x_t + e_t
    \end{gather*}
    In Panel A, $y_t$ is one of StockGPT-based portfolios and $x_t$ are short-term reversal (ST\_Rev), momentum (Mom), long-term reversal (LT\_Rev), market (MKT), value (HML), size (SMB), profitability (RMW), and investment (CMA) from \citet{fama2015five}, as well as investment (R\_IA), return on equity (R\_ROE), and earnings growth (R\_EG) from \citet{hou2021augmented}. In Panel B, $y_t$ is one of the factors and $x_t$ is one of StockGPT-based portfolios. $\alpha$ is in annualized percentage points and $R^2$ is in percentage points. $t_{\beta}$ is computed with Newey-West standard error using 20 lags. The sample is daily from January 2001 to December 2023.
    \vspace{5pt}
    
    \centering
    \vspace{5pt}
    Panel A: Stock Factors Span StockGPT
    \vspace{5pt}
    
    \begin{adjustbox}{width = 0.9\columnwidth, center}
        
  \providecommand{\huxb}[2]{\arrayrulecolor[RGB]{#1}\global\arrayrulewidth=#2pt}
  \providecommand{\huxvb}[2]{\color[RGB]{#1}\vrule width #2pt}
  \providecommand{\huxtpad}[1]{\rule{0pt}{#1}}
  \providecommand{\huxbpad}[1]{\rule[-#1]{0pt}{#1}}
\begin{tabular}{l l l l l l l l l l}

\hhline{>{\huxb{0, 0, 0}{0.8}}->{\huxb{0, 0, 0}{0.8}}->{\huxb{0, 0, 0}{0.8}}->{\huxb{0, 0, 0}{0.8}}->{\huxb{0, 0, 0}{0.8}}->{\huxb{0, 0, 0}{0.8}}->{\huxb{0, 0, 0}{0.8}}->{\huxb{0, 0, 0}{0.8}}->{\huxb{0, 0, 0}{0.8}}->{\huxb{0, 0, 0}{0.8}}-}
\arrayrulecolor{black}

\multicolumn{1}{!{\huxvb{0, 0, 0}{0}}l!{\huxvb{0, 0, 0}{0}}}{\huxtpad{0pt + 1em}\raggedright \hspace{0pt}  \hspace{0pt}\huxbpad{0pt}} &
\multicolumn{1}{l!{\huxvb{0, 0, 0}{0}}}{\huxtpad{0pt + 1em}\raggedright \hspace{0pt}  \hspace{0pt}\huxbpad{0pt}} &
\multicolumn{1}{l!{\huxvb{0, 0, 0}{0}}}{\huxtpad{0pt + 1em}\raggedright \hspace{0pt} $\beta$ \hspace{0pt}\huxbpad{0pt}} &
\multicolumn{1}{l!{\huxvb{0, 0, 0}{0}}}{\huxtpad{0pt + 1em}\raggedright \hspace{0pt} $t_{\beta}$ \hspace{0pt}\huxbpad{0pt}} &
\multicolumn{1}{l!{\huxvb{0, 0, 0}{0}}}{\huxtpad{0pt + 1em}\raggedright \hspace{0pt} $\beta$ \hspace{0pt}\huxbpad{0pt}} &
\multicolumn{1}{l!{\huxvb{0, 0, 0}{0}}}{\huxtpad{0pt + 1em}\raggedright \hspace{0pt} $t_{\beta}$ \hspace{0pt}\huxbpad{0pt}} &
\multicolumn{1}{l!{\huxvb{0, 0, 0}{0}}}{\huxtpad{0pt + 1em}\raggedright \hspace{0pt} $\beta$ \hspace{0pt}\huxbpad{0pt}} &
\multicolumn{1}{l!{\huxvb{0, 0, 0}{0}}}{\huxtpad{0pt + 1em}\raggedright \hspace{0pt} $t_{\beta}$ \hspace{0pt}\huxbpad{0pt}} &
\multicolumn{1}{l!{\huxvb{0, 0, 0}{0}}}{\huxtpad{0pt + 1em}\raggedright \hspace{0pt} $\beta$ \hspace{0pt}\huxbpad{0pt}} &
\multicolumn{1}{l!{\huxvb{0, 0, 0}{0}}}{\huxtpad{0pt + 1em}\raggedright \hspace{0pt} $t_{\beta}$ \hspace{0pt}\huxbpad{0pt}} \tabularnewline
\end{tabular}
    \end{adjustbox}

    \vspace{5pt}
    Panel B: StockGPT Spans Stock Factors
    \vspace{5pt}
    
    \begin{adjustbox}{width = \columnwidth, center}
        
  \providecommand{\huxb}[2]{\arrayrulecolor[RGB]{#1}\global\arrayrulewidth=#2pt}
  \providecommand{\huxvb}[2]{\color[RGB]{#1}\vrule width #2pt}
  \providecommand{\huxtpad}[1]{\rule{0pt}{#1}}
  \providecommand{\huxbpad}[1]{\rule[-#1]{0pt}{#1}}
\begin{tabular}{l l l l l l l l l l l l l}

\hhline{>{\huxb{0, 0, 0}{0.8}}->{\huxb{0, 0, 0}{0.8}}->{\huxb{0, 0, 0}{0.8}}->{\huxb{0, 0, 0}{0.8}}->{\huxb{0, 0, 0}{0.8}}->{\huxb{0, 0, 0}{0.8}}->{\huxb{0, 0, 0}{0.8}}->{\huxb{0, 0, 0}{0.8}}->{\huxb{0, 0, 0}{0.8}}->{\huxb{0, 0, 0}{0.8}}->{\huxb{0, 0, 0}{0.8}}->{\huxb{0, 0, 0}{0.8}}->{\huxb{0, 0, 0}{0.8}}-}
\arrayrulecolor{black}

\multicolumn{1}{!{\huxvb{0, 0, 0}{0}}l!{\huxvb{0, 0, 0}{0}}}{\huxtpad{0pt + 1em}\raggedright \hspace{0pt}  \hspace{0pt}\huxbpad{0pt}} &
\multicolumn{1}{l!{\huxvb{0, 0, 0}{0}}}{\huxtpad{0pt + 1em}\raggedright \hspace{0pt}  \hspace{0pt}\huxbpad{0pt}} &
\multicolumn{1}{r!{\huxvb{0, 0, 0}{0}}}{\huxtpad{0pt + 1em}\raggedleft \hspace{0pt} ST\_Rev \hspace{0pt}\huxbpad{0pt}} &
\multicolumn{1}{r!{\huxvb{0, 0, 0}{0}}}{\huxtpad{0pt + 1em}\raggedleft \hspace{0pt} Mom \hspace{0pt}\huxbpad{0pt}} &
\multicolumn{1}{r!{\huxvb{0, 0, 0}{0}}}{\huxtpad{0pt + 1em}\raggedleft \hspace{0pt} LT\_Rev \hspace{0pt}\huxbpad{0pt}} &
\multicolumn{1}{r!{\huxvb{0, 0, 0}{0}}}{\huxtpad{0pt + 1em}\raggedleft \hspace{0pt} MKT \hspace{0pt}\huxbpad{0pt}} &
\multicolumn{1}{r!{\huxvb{0, 0, 0}{0}}}{\huxtpad{0pt + 1em}\raggedleft \hspace{0pt} HML \hspace{0pt}\huxbpad{0pt}} &
\multicolumn{1}{r!{\huxvb{0, 0, 0}{0}}}{\huxtpad{0pt + 1em}\raggedleft \hspace{0pt} SMB \hspace{0pt}\huxbpad{0pt}} &
\multicolumn{1}{r!{\huxvb{0, 0, 0}{0}}}{\huxtpad{0pt + 1em}\raggedleft \hspace{0pt} RMW \hspace{0pt}\huxbpad{0pt}} &
\multicolumn{1}{r!{\huxvb{0, 0, 0}{0}}}{\huxtpad{0pt + 1em}\raggedleft \hspace{0pt} CMA \hspace{0pt}\huxbpad{0pt}} &
\multicolumn{1}{r!{\huxvb{0, 0, 0}{0}}}{\huxtpad{0pt + 1em}\raggedleft \hspace{0pt} R\_IA \hspace{0pt}\huxbpad{0pt}} &
\multicolumn{1}{r!{\huxvb{0, 0, 0}{0}}}{\huxtpad{0pt + 1em}\raggedleft \hspace{0pt} R\_ROE \hspace{0pt}\huxbpad{0pt}} &
\multicolumn{1}{r!{\huxvb{0, 0, 0}{0}}}{\huxtpad{0pt + 1em}\raggedleft \hspace{0pt} R\_EG \hspace{0pt}\huxbpad{0pt}} \tabularnewline
\end{tabular}
    \end{adjustbox}
    
    \label{tab:alpha}
\end{table}

\begin{table}[t]
    \caption{Monthly Fama-MacBeth Regression}

    \vspace{5pt}
    \footnotesize 
    \noindent This table reports the time series averages of slopes and adjusted $R^2$'s of the following cross-sectional regression
    \begin{gather*}
    x_{it+1} = a_t + b_t \times \hat{x}_{it+1} + e_{it+1}
    \end{gather*}
    where $x_{it+1}$ is the actual realized return of stock $i$ in month $t+1$ and $\hat{x}_{it+1}$ is its StockGPT return forecast. Returns are in basis points and $R^2$ in percentage points. $t$ is the $t$-statistic of time-series mean of $b_t$ computed using \citet{newey1987hypothesis} standard error with 4 lags. Horizon 1 (2) means comparing return forecasts for month $t+1$ with actual returns in month $t+1$ ($t+2$). The sample is monthly from February 2001 to December 2023.
    \vspace{5pt}
    
    \centering
    
  \providecommand{\huxb}[2]{\arrayrulecolor[RGB]{#1}\global\arrayrulewidth=#2pt}
  \providecommand{\huxvb}[2]{\color[RGB]{#1}\vrule width #2pt}
  \providecommand{\huxtpad}[1]{\rule{0pt}{#1}}
  \providecommand{\huxbpad}[1]{\rule[-#1]{0pt}{#1}}
\begin{tabular}{l l l l}

\hhline{>{\huxb{0, 0, 0}{0.8}}->{\huxb{0, 0, 0}{0.8}}->{\huxb{0, 0, 0}{0.8}}->{\huxb{0, 0, 0}{0.8}}-}
\arrayrulecolor{black}

\multicolumn{1}{!{\huxvb{0, 0, 0}{0}}c!{\huxvb{0, 0, 0}{0}}}{\huxtpad{0pt + 1em}\centering \hspace{6pt} Horizon \hspace{6pt}\huxbpad{0pt}} &
\multicolumn{1}{c!{\huxvb{0, 0, 0}{0}}}{\huxtpad{0pt + 1em}\centering \hspace{6pt} b \hspace{6pt}\huxbpad{0pt}} &
\multicolumn{1}{c!{\huxvb{0, 0, 0}{0}}}{\huxtpad{0pt + 1em}\centering \hspace{6pt} $t$ \hspace{6pt}\huxbpad{0pt}} &
\multicolumn{1}{c!{\huxvb{0, 0, 0}{0}}}{\huxtpad{0pt + 1em}\centering \hspace{6pt} $R^2$ \hspace{6pt}\huxbpad{0pt}} \tabularnewline[-0.5pt]

\hhline{>{\huxb{0, 0, 0}{0.8}}->{\huxb{0, 0, 0}{0.8}}->{\huxb{0, 0, 0}{0.8}}->{\huxb{0, 0, 0}{0.8}}-}
\arrayrulecolor{black}

\multicolumn{1}{!{\huxvb{0, 0, 0}{0}}c!{\huxvb{0, 0, 0}{0}}}{\huxtpad{0pt + 1em}\centering \hspace{6pt} 1 \hspace{6pt}\huxbpad{0pt}} &
\multicolumn{1}{c!{\huxvb{0, 0, 0}{0}}}{\huxtpad{0pt + 1em}\centering \hspace{6pt} 3.01 \hspace{6pt}\huxbpad{0pt}} &
\multicolumn{1}{c!{\huxvb{0, 0, 0}{0}}}{\huxtpad{0pt + 1em}\centering \hspace{6pt} 2.49 \hspace{6pt}\huxbpad{0pt}} &
\multicolumn{1}{c!{\huxvb{0, 0, 0}{0}}}{\huxtpad{0pt + 1em}\centering \hspace{6pt} 0.55 \hspace{6pt}\huxbpad{0pt}} \tabularnewline[-0.5pt]

\hhline{}
\arrayrulecolor{black}

\multicolumn{1}{!{\huxvb{0, 0, 0}{0}}c!{\huxvb{0, 0, 0}{0}}}{\huxtpad{0pt + 1em}\centering \hspace{6pt} 2 \hspace{6pt}\huxbpad{0pt}} &
\multicolumn{1}{c!{\huxvb{0, 0, 0}{0}}}{\huxtpad{0pt + 1em}\centering \hspace{6pt} -0.08 \hspace{6pt}\huxbpad{0pt}} &
\multicolumn{1}{c!{\huxvb{0, 0, 0}{0}}}{\huxtpad{0pt + 1em}\centering \hspace{6pt} -0.08 \hspace{6pt}\huxbpad{0pt}} &
\multicolumn{1}{c!{\huxvb{0, 0, 0}{0}}}{\huxtpad{0pt + 1em}\centering \hspace{6pt} 0.43 \hspace{6pt}\huxbpad{0pt}} \tabularnewline[-0.5pt]

\hhline{>{\huxb{0, 0, 0}{0.8}}->{\huxb{0, 0, 0}{0.8}}->{\huxb{0, 0, 0}{0.8}}->{\huxb{0, 0, 0}{0.8}}-}
\arrayrulecolor{black}
\end{tabular}
    \label{tab:monthly_fmb}
\end{table}

\begin{table}[t]
    \caption{Monthly Portfolio Statistics}

    \vspace{5pt}
    \footnotesize
    \noindent Panel A reports the return statistics of the monthly equal-weighted long-short StockGPT-based portfolios. Mcap Filter refers to the monthly market cap percentile under which stocks are removed and Price Filter refers to the price level under which stocks are removed. Panel B reports the return statistics of stock factors including short-term reversal (ST\_Rev), momentum (Mom), long-term reversal (LT\_Rev), market (MKT), value (HML), size (SMB), profitability (RMW), and investment (CMA) from \citet{fama2015five}, as well as investment (R\_IA), return on equity (R\_ROE), and earnings growth (R\_EG) from \citet{hou2021augmented}. Mean and SD (standard deviation) are in annualized percentage points; Mean/SD (Sharpe ratio) is annualized; Min, Max, and MDD (max drawdown) are in percentage points; and t-Mean is $t$-statistic of the mean portfolio return using Newey-West standard error with 4 lags. The sample is monthly from February 2001 to December 2023.
    \vspace{5pt}
    
    \centering
    \vspace{5pt}
    Panel A: StockGPT Portfolios
    \vspace{5pt}
    
  \providecommand{\huxb}[2]{\arrayrulecolor[RGB]{#1}\global\arrayrulewidth=#2pt}
  \providecommand{\huxvb}[2]{\color[RGB]{#1}\vrule width #2pt}
  \providecommand{\huxtpad}[1]{\rule{0pt}{#1}}
  \providecommand{\huxbpad}[1]{\rule[-#1]{0pt}{#1}}
\begin{tabular}{l l l l l l l l l l}

\hhline{>{\huxb{0, 0, 0}{0.8}}->{\huxb{0, 0, 0}{0.8}}->{\huxb{0, 0, 0}{0.8}}->{\huxb{0, 0, 0}{0.8}}->{\huxb{0, 0, 0}{0.8}}->{\huxb{0, 0, 0}{0.8}}->{\huxb{0, 0, 0}{0.8}}->{\huxb{0, 0, 0}{0.8}}->{\huxb{0, 0, 0}{0.8}}->{\huxb{0, 0, 0}{0.8}}-}
\arrayrulecolor{black}

\multicolumn{1}{!{\huxvb{0, 0, 0}{0}}c!{\huxvb{0, 0, 0}{0}}}{\huxtpad{0pt + 1em}\centering \hspace{0pt} Exchange \hspace{0pt}\huxbpad{0pt}} &
\multicolumn{1}{c!{\huxvb{0, 0, 0}{0}}}{\huxtpad{0pt + 1em}\centering \hspace{0pt} MCap Filter \hspace{0pt}\huxbpad{0pt}} &
\multicolumn{1}{c!{\huxvb{0, 0, 0}{0}}}{\huxtpad{0pt + 1em}\centering \hspace{0pt} Price Filter \hspace{0pt}\huxbpad{0pt}} &
\multicolumn{1}{c!{\huxvb{0, 0, 0}{0}}}{\huxtpad{0pt + 1em}\centering \hspace{0pt} Mean \hspace{0pt}\huxbpad{0pt}} &
\multicolumn{1}{c!{\huxvb{0, 0, 0}{0}}}{\huxtpad{0pt + 1em}\centering \hspace{0pt} t-Mean \hspace{0pt}\huxbpad{0pt}} &
\multicolumn{1}{c!{\huxvb{0, 0, 0}{0}}}{\huxtpad{0pt + 1em}\centering \hspace{0pt} SD \hspace{0pt}\huxbpad{0pt}} &
\multicolumn{1}{c!{\huxvb{0, 0, 0}{0}}}{\huxtpad{0pt + 1em}\centering \hspace{0pt} Mean/SD \hspace{0pt}\huxbpad{0pt}} &
\multicolumn{1}{c!{\huxvb{0, 0, 0}{0}}}{\huxtpad{0pt + 1em}\centering \hspace{0pt} Min \hspace{0pt}\huxbpad{0pt}} &
\multicolumn{1}{c!{\huxvb{0, 0, 0}{0}}}{\huxtpad{0pt + 1em}\centering \hspace{0pt} Max \hspace{0pt}\huxbpad{0pt}} &
\multicolumn{1}{c!{\huxvb{0, 0, 0}{0}}}{\huxtpad{0pt + 1em}\centering \hspace{0pt} MDD \hspace{0pt}\huxbpad{0pt}} \tabularnewline
\end{tabular}

    \vspace{5pt}
    Panel B: Stock Factors
    \vspace{5pt}
    
  \providecommand{\huxb}[2]{\arrayrulecolor[RGB]{#1}\global\arrayrulewidth=#2pt}
  \providecommand{\huxvb}[2]{\color[RGB]{#1}\vrule width #2pt}
  \providecommand{\huxtpad}[1]{\rule{0pt}{#1}}
  \providecommand{\huxbpad}[1]{\rule[-#1]{0pt}{#1}}
\begin{tabular}{l l l l l l l l}

\hhline{>{\huxb{0, 0, 0}{0.8}}->{\huxb{0, 0, 0}{0.8}}->{\huxb{0, 0, 0}{0.8}}->{\huxb{0, 0, 0}{0.8}}->{\huxb{0, 0, 0}{0.8}}->{\huxb{0, 0, 0}{0.8}}->{\huxb{0, 0, 0}{0.8}}->{\huxb{0, 0, 0}{0.8}}-}
\arrayrulecolor{black}

\multicolumn{1}{!{\huxvb{0, 0, 0}{0}}c!{\huxvb{0, 0, 0}{0}}}{\huxtpad{0pt + 1em}\centering \hspace{0pt} Factor \hspace{0pt}\huxbpad{0pt}} &
\multicolumn{1}{c!{\huxvb{0, 0, 0}{0}}}{\huxtpad{0pt + 1em}\centering \hspace{0pt} Mean \hspace{0pt}\huxbpad{0pt}} &
\multicolumn{1}{c!{\huxvb{0, 0, 0}{0}}}{\huxtpad{0pt + 1em}\centering \hspace{0pt} t-Mean \hspace{0pt}\huxbpad{0pt}} &
\multicolumn{1}{c!{\huxvb{0, 0, 0}{0}}}{\huxtpad{0pt + 1em}\centering \hspace{0pt} SD \hspace{0pt}\huxbpad{0pt}} &
\multicolumn{1}{c!{\huxvb{0, 0, 0}{0}}}{\huxtpad{0pt + 1em}\centering \hspace{0pt} Mean/SD \hspace{0pt}\huxbpad{0pt}} &
\multicolumn{1}{c!{\huxvb{0, 0, 0}{0}}}{\huxtpad{0pt + 1em}\centering \hspace{0pt} Min \hspace{0pt}\huxbpad{0pt}} &
\multicolumn{1}{c!{\huxvb{0, 0, 0}{0}}}{\huxtpad{0pt + 1em}\centering \hspace{0pt} Max \hspace{0pt}\huxbpad{0pt}} &
\multicolumn{1}{c!{\huxvb{0, 0, 0}{0}}}{\huxtpad{0pt + 1em}\centering \hspace{0pt} MDD \hspace{0pt}\huxbpad{0pt}} \tabularnewline
\end{tabular}

    \label{tab:monthly_port}
\end{table}

\begin{table}[t]
    \caption{Monthly Spanning Tests}

    \vspace{5pt}
    \footnotesize
    \noindent This table reports results of the following spanning test
    \begin{gather*}
        y_t = \alpha + \beta \times x_t + e_t
    \end{gather*}
    In Panel A, $y_t$ is the equal-weighted monthly-rebalanced StockGPT-based portfolio and $x_t$ are short-term reversal (ST\_Rev), momentum (Mom), long-term reversal (LT\_Rev), market (MKT), value (HML), size (SMB), profitability (RMW), and investment (CMA) from \citet{fama2015five}, as well as investment (R\_IA), return on equity (R\_ROE), and earnings growth (R\_EG) from \citet{hou2021augmented}. In Panel B, $y_t$ is one of the factors and $x_t$ is the StockGPT-based portfolio. $\alpha$ is in annualized percentage points and $R^2$ is in percentage points. $t_{\beta}$ is computed with Newey-West standard error using 4 lags. The sample is monthly from February 2001 to December 2023.
    \vspace{5pt}
    
    \centering
    \vspace{5pt}
    Panel A: Stock Factors Span StockGPT
    \vspace{5pt}
    
    \begin{adjustbox}{width = 0.9\columnwidth, center}
        
  \providecommand{\huxb}[2]{\arrayrulecolor[RGB]{#1}\global\arrayrulewidth=#2pt}
  \providecommand{\huxvb}[2]{\color[RGB]{#1}\vrule width #2pt}
  \providecommand{\huxtpad}[1]{\rule{0pt}{#1}}
  \providecommand{\huxbpad}[1]{\rule[-#1]{0pt}{#1}}
\begin{tabular}{l l l l l l l l l}

\hhline{>{\huxb{0, 0, 0}{0.8}}->{\huxb{0, 0, 0}{0.8}}->{\huxb{0, 0, 0}{0.8}}->{\huxb{0, 0, 0}{0.8}}->{\huxb{0, 0, 0}{0.8}}->{\huxb{0, 0, 0}{0.8}}->{\huxb{0, 0, 0}{0.8}}->{\huxb{0, 0, 0}{0.8}}->{\huxb{0, 0, 0}{0.8}}-}
\arrayrulecolor{black}

\multicolumn{1}{!{\huxvb{0, 0, 0}{0}}l!{\huxvb{0, 0, 0}{0}}}{\huxtpad{0pt + 1em}\raggedright \hspace{0pt}  \hspace{0pt}\huxbpad{0pt}} &
\multicolumn{1}{l!{\huxvb{0, 0, 0}{0}}}{\huxtpad{0pt + 1em}\raggedright \hspace{0pt} $\beta$ \hspace{0pt}\huxbpad{0pt}} &
\multicolumn{1}{l!{\huxvb{0, 0, 0}{0}}}{\huxtpad{0pt + 1em}\raggedright \hspace{0pt} $t_{\beta}$ \hspace{0pt}\huxbpad{0pt}} &
\multicolumn{1}{l!{\huxvb{0, 0, 0}{0}}}{\huxtpad{0pt + 1em}\raggedright \hspace{0pt} $\beta$ \hspace{0pt}\huxbpad{0pt}} &
\multicolumn{1}{l!{\huxvb{0, 0, 0}{0}}}{\huxtpad{0pt + 1em}\raggedright \hspace{0pt} $t_{\beta}$ \hspace{0pt}\huxbpad{0pt}} &
\multicolumn{1}{l!{\huxvb{0, 0, 0}{0}}}{\huxtpad{0pt + 1em}\raggedright \hspace{0pt} $\beta$ \hspace{0pt}\huxbpad{0pt}} &
\multicolumn{1}{l!{\huxvb{0, 0, 0}{0}}}{\huxtpad{0pt + 1em}\raggedright \hspace{0pt} $t_{\beta}$ \hspace{0pt}\huxbpad{0pt}} &
\multicolumn{1}{l!{\huxvb{0, 0, 0}{0}}}{\huxtpad{0pt + 1em}\raggedright \hspace{0pt} $\beta$ \hspace{0pt}\huxbpad{0pt}} &
\multicolumn{1}{l!{\huxvb{0, 0, 0}{0}}}{\huxtpad{0pt + 1em}\raggedright \hspace{0pt} $t_{\beta}$ \hspace{0pt}\huxbpad{0pt}} \tabularnewline[-0.5pt]

\hhline{>{\huxb{0, 0, 0}{0.8}}->{\huxb{0, 0, 0}{0.8}}->{\huxb{0, 0, 0}{0.8}}->{\huxb{0, 0, 0}{0.8}}->{\huxb{0, 0, 0}{0.8}}->{\huxb{0, 0, 0}{0.8}}->{\huxb{0, 0, 0}{0.8}}->{\huxb{0, 0, 0}{0.8}}->{\huxb{0, 0, 0}{0.8}}-}
\arrayrulecolor{black}

\multicolumn{1}{!{\huxvb{0, 0, 0}{0}}l!{\huxvb{0, 0, 0}{0}}}{\huxtpad{0pt + 1em}\raggedright \hspace{0pt} $\alpha$ \hspace{0pt}\huxbpad{0pt}} &
\multicolumn{1}{r!{\huxvb{0, 0, 0}{0}}}{\huxtpad{0pt + 1em}\raggedleft \hspace{0pt} 15.58 \hspace{0pt}\huxbpad{0pt}} &
\multicolumn{1}{r!{\huxvb{0, 0, 0}{0}}}{\huxtpad{0pt + 1em}\raggedleft \hspace{0pt} 4.66 \hspace{0pt}\huxbpad{0pt}} &
\multicolumn{1}{r!{\huxvb{0, 0, 0}{0}}}{\huxtpad{0pt + 1em}\raggedleft \hspace{0pt} 14.94 \hspace{0pt}\huxbpad{0pt}} &
\multicolumn{1}{r!{\huxvb{0, 0, 0}{0}}}{\huxtpad{0pt + 1em}\raggedleft \hspace{0pt} 4.16 \hspace{0pt}\huxbpad{0pt}} &
\multicolumn{1}{r!{\huxvb{0, 0, 0}{0}}}{\huxtpad{0pt + 1em}\raggedleft \hspace{0pt} 15.43 \hspace{0pt}\huxbpad{0pt}} &
\multicolumn{1}{r!{\huxvb{0, 0, 0}{0}}}{\huxtpad{0pt + 1em}\raggedleft \hspace{0pt} 4.79 \hspace{0pt}\huxbpad{0pt}} &
\multicolumn{1}{r!{\huxvb{0, 0, 0}{0}}}{\huxtpad{0pt + 1em}\raggedleft \hspace{0pt} 15.27 \hspace{0pt}\huxbpad{0pt}} &
\multicolumn{1}{r!{\huxvb{0, 0, 0}{0}}}{\huxtpad{0pt + 1em}\raggedleft \hspace{0pt} 4.87 \hspace{0pt}\huxbpad{0pt}} \tabularnewline[-0.5pt]

\hhline{}
\arrayrulecolor{black}

\multicolumn{1}{!{\huxvb{0, 0, 0}{0}}l!{\huxvb{0, 0, 0}{0}}}{\huxtpad{0pt + 1em}\raggedright \hspace{0pt} ST\_Rev \hspace{0pt}\huxbpad{0pt}} &
\multicolumn{1}{r!{\huxvb{0, 0, 0}{0}}}{\huxtpad{0pt + 1em}\raggedleft \hspace{0pt} -0.06 \hspace{0pt}\huxbpad{0pt}} &
\multicolumn{1}{r!{\huxvb{0, 0, 0}{0}}}{\huxtpad{0pt + 1em}\raggedleft \hspace{0pt} -0.66 \hspace{0pt}\huxbpad{0pt}} &
\multicolumn{1}{r!{\huxvb{0, 0, 0}{0}}}{\huxtpad{0pt + 1em}\raggedleft \hspace{0pt} -0.08 \hspace{0pt}\huxbpad{0pt}} &
\multicolumn{1}{r!{\huxvb{0, 0, 0}{0}}}{\huxtpad{0pt + 1em}\raggedleft \hspace{0pt} -0.77 \hspace{0pt}\huxbpad{0pt}} &
\multicolumn{1}{r!{\huxvb{0, 0, 0}{0}}}{\huxtpad{0pt + 1em}\raggedleft \hspace{0pt} \hphantom{0}\hphantom{0}\hphantom{0} \hspace{0pt}\huxbpad{0pt}} &
\multicolumn{1}{r!{\huxvb{0, 0, 0}{0}}}{\huxtpad{0pt + 1em}\raggedleft \hspace{0pt} \hphantom{0}\hphantom{0}\hphantom{0} \hspace{0pt}\huxbpad{0pt}} &
\multicolumn{1}{r!{\huxvb{0, 0, 0}{0}}}{\huxtpad{0pt + 1em}\raggedleft \hspace{0pt} \hphantom{0}\hphantom{0}\hphantom{0} \hspace{0pt}\huxbpad{0pt}} &
\multicolumn{1}{r!{\huxvb{0, 0, 0}{0}}}{\huxtpad{0pt + 1em}\raggedleft \hspace{0pt} \hphantom{0}\hphantom{0}\hphantom{0} \hspace{0pt}\huxbpad{0pt}} \tabularnewline[-0.5pt]

\hhline{}
\arrayrulecolor{black}

\multicolumn{1}{!{\huxvb{0, 0, 0}{0}}l!{\huxvb{0, 0, 0}{0}}}{\huxtpad{0pt + 1em}\raggedright \hspace{0pt} Mom \hspace{0pt}\huxbpad{0pt}} &
\multicolumn{1}{r!{\huxvb{0, 0, 0}{0}}}{\huxtpad{0pt + 1em}\raggedleft \hspace{0pt} -0.21 \hspace{0pt}\huxbpad{0pt}} &
\multicolumn{1}{r!{\huxvb{0, 0, 0}{0}}}{\huxtpad{0pt + 1em}\raggedleft \hspace{0pt} -1.82 \hspace{0pt}\huxbpad{0pt}} &
\multicolumn{1}{r!{\huxvb{0, 0, 0}{0}}}{\huxtpad{0pt + 1em}\raggedleft \hspace{0pt} -0.25 \hspace{0pt}\huxbpad{0pt}} &
\multicolumn{1}{r!{\huxvb{0, 0, 0}{0}}}{\huxtpad{0pt + 1em}\raggedleft \hspace{0pt} -1.99 \hspace{0pt}\huxbpad{0pt}} &
\multicolumn{1}{r!{\huxvb{0, 0, 0}{0}}}{\huxtpad{0pt + 1em}\raggedleft \hspace{0pt} \hphantom{0}\hphantom{0}\hphantom{0} \hspace{0pt}\huxbpad{0pt}} &
\multicolumn{1}{r!{\huxvb{0, 0, 0}{0}}}{\huxtpad{0pt + 1em}\raggedleft \hspace{0pt} \hphantom{0}\hphantom{0}\hphantom{0} \hspace{0pt}\huxbpad{0pt}} &
\multicolumn{1}{r!{\huxvb{0, 0, 0}{0}}}{\huxtpad{0pt + 1em}\raggedleft \hspace{0pt} \hphantom{0}\hphantom{0}\hphantom{0} \hspace{0pt}\huxbpad{0pt}} &
\multicolumn{1}{r!{\huxvb{0, 0, 0}{0}}}{\huxtpad{0pt + 1em}\raggedleft \hspace{0pt} \hphantom{0}\hphantom{0}\hphantom{0} \hspace{0pt}\huxbpad{0pt}} \tabularnewline[-0.5pt]

\hhline{}
\arrayrulecolor{black}

\multicolumn{1}{!{\huxvb{0, 0, 0}{0}}l!{\huxvb{0, 0, 0}{0}}}{\huxtpad{0pt + 1em}\raggedright \hspace{0pt} LT\_Rev \hspace{0pt}\huxbpad{0pt}} &
\multicolumn{1}{r!{\huxvb{0, 0, 0}{0}}}{\huxtpad{0pt + 1em}\raggedleft \hspace{0pt} -0.49 \hspace{0pt}\huxbpad{0pt}} &
\multicolumn{1}{r!{\huxvb{0, 0, 0}{0}}}{\huxtpad{0pt + 1em}\raggedleft \hspace{0pt} -3.25 \hspace{0pt}\huxbpad{0pt}} &
\multicolumn{1}{r!{\huxvb{0, 0, 0}{0}}}{\huxtpad{0pt + 1em}\raggedleft \hspace{0pt} -0.31 \hspace{0pt}\huxbpad{0pt}} &
\multicolumn{1}{r!{\huxvb{0, 0, 0}{0}}}{\huxtpad{0pt + 1em}\raggedleft \hspace{0pt} -2.71 \hspace{0pt}\huxbpad{0pt}} &
\multicolumn{1}{r!{\huxvb{0, 0, 0}{0}}}{\huxtpad{0pt + 1em}\raggedleft \hspace{0pt} \hphantom{0}\hphantom{0}\hphantom{0} \hspace{0pt}\huxbpad{0pt}} &
\multicolumn{1}{r!{\huxvb{0, 0, 0}{0}}}{\huxtpad{0pt + 1em}\raggedleft \hspace{0pt} \hphantom{0}\hphantom{0}\hphantom{0} \hspace{0pt}\huxbpad{0pt}} &
\multicolumn{1}{r!{\huxvb{0, 0, 0}{0}}}{\huxtpad{0pt + 1em}\raggedleft \hspace{0pt} \hphantom{0}\hphantom{0}\hphantom{0} \hspace{0pt}\huxbpad{0pt}} &
\multicolumn{1}{r!{\huxvb{0, 0, 0}{0}}}{\huxtpad{0pt + 1em}\raggedleft \hspace{0pt} \hphantom{0}\hphantom{0}\hphantom{0} \hspace{0pt}\huxbpad{0pt}} \tabularnewline[-0.5pt]

\hhline{}
\arrayrulecolor{black}

\multicolumn{1}{!{\huxvb{0, 0, 0}{0}}l!{\huxvb{0, 0, 0}{0}}}{\huxtpad{0pt + 1em}\raggedright \hspace{0pt} MKT \hspace{0pt}\huxbpad{0pt}} &
\multicolumn{1}{r!{\huxvb{0, 0, 0}{0}}}{\huxtpad{0pt + 1em}\raggedleft \hspace{0pt} 0.04 \hspace{0pt}\huxbpad{0pt}} &
\multicolumn{1}{r!{\huxvb{0, 0, 0}{0}}}{\huxtpad{0pt + 1em}\raggedleft \hspace{0pt} 0.70 \hspace{0pt}\huxbpad{0pt}} &
\multicolumn{1}{r!{\huxvb{0, 0, 0}{0}}}{\huxtpad{0pt + 1em}\raggedleft \hspace{0pt} \hphantom{0}\hphantom{0}\hphantom{0} \hspace{0pt}\huxbpad{0pt}} &
\multicolumn{1}{r!{\huxvb{0, 0, 0}{0}}}{\huxtpad{0pt + 1em}\raggedleft \hspace{0pt} \hphantom{0}\hphantom{0}\hphantom{0} \hspace{0pt}\huxbpad{0pt}} &
\multicolumn{1}{r!{\huxvb{0, 0, 0}{0}}}{\huxtpad{0pt + 1em}\raggedleft \hspace{0pt} 0.08 \hspace{0pt}\huxbpad{0pt}} &
\multicolumn{1}{r!{\huxvb{0, 0, 0}{0}}}{\huxtpad{0pt + 1em}\raggedleft \hspace{0pt} 1.10 \hspace{0pt}\huxbpad{0pt}} &
\multicolumn{1}{r!{\huxvb{0, 0, 0}{0}}}{\huxtpad{0pt + 1em}\raggedleft \hspace{0pt} 0.06 \hspace{0pt}\huxbpad{0pt}} &
\multicolumn{1}{r!{\huxvb{0, 0, 0}{0}}}{\huxtpad{0pt + 1em}\raggedleft \hspace{0pt} 0.84 \hspace{0pt}\huxbpad{0pt}} \tabularnewline[-0.5pt]

\hhline{}
\arrayrulecolor{black}

\multicolumn{1}{!{\huxvb{0, 0, 0}{0}}l!{\huxvb{0, 0, 0}{0}}}{\huxtpad{0pt + 1em}\raggedright \hspace{0pt} HML \hspace{0pt}\huxbpad{0pt}} &
\multicolumn{1}{r!{\huxvb{0, 0, 0}{0}}}{\huxtpad{0pt + 1em}\raggedleft \hspace{0pt} 0.09 \hspace{0pt}\huxbpad{0pt}} &
\multicolumn{1}{r!{\huxvb{0, 0, 0}{0}}}{\huxtpad{0pt + 1em}\raggedleft \hspace{0pt} 0.51 \hspace{0pt}\huxbpad{0pt}} &
\multicolumn{1}{r!{\huxvb{0, 0, 0}{0}}}{\huxtpad{0pt + 1em}\raggedleft \hspace{0pt} \hphantom{0}\hphantom{0}\hphantom{0} \hspace{0pt}\huxbpad{0pt}} &
\multicolumn{1}{r!{\huxvb{0, 0, 0}{0}}}{\huxtpad{0pt + 1em}\raggedleft \hspace{0pt} \hphantom{0}\hphantom{0}\hphantom{0} \hspace{0pt}\huxbpad{0pt}} &
\multicolumn{1}{r!{\huxvb{0, 0, 0}{0}}}{\huxtpad{0pt + 1em}\raggedleft \hspace{0pt} 0.01 \hspace{0pt}\huxbpad{0pt}} &
\multicolumn{1}{r!{\huxvb{0, 0, 0}{0}}}{\huxtpad{0pt + 1em}\raggedleft \hspace{0pt} 0.10 \hspace{0pt}\huxbpad{0pt}} &
\multicolumn{1}{r!{\huxvb{0, 0, 0}{0}}}{\huxtpad{0pt + 1em}\raggedleft \hspace{0pt} \hphantom{0}\hphantom{0}\hphantom{0} \hspace{0pt}\huxbpad{0pt}} &
\multicolumn{1}{r!{\huxvb{0, 0, 0}{0}}}{\huxtpad{0pt + 1em}\raggedleft \hspace{0pt} \hphantom{0}\hphantom{0}\hphantom{0} \hspace{0pt}\huxbpad{0pt}} \tabularnewline[-0.5pt]

\hhline{}
\arrayrulecolor{black}

\multicolumn{1}{!{\huxvb{0, 0, 0}{0}}l!{\huxvb{0, 0, 0}{0}}}{\huxtpad{0pt + 1em}\raggedright \hspace{0pt} SMB \hspace{0pt}\huxbpad{0pt}} &
\multicolumn{1}{r!{\huxvb{0, 0, 0}{0}}}{\huxtpad{0pt + 1em}\raggedleft \hspace{0pt} -0.18 \hspace{0pt}\huxbpad{0pt}} &
\multicolumn{1}{r!{\huxvb{0, 0, 0}{0}}}{\huxtpad{0pt + 1em}\raggedleft \hspace{0pt} -1.54 \hspace{0pt}\huxbpad{0pt}} &
\multicolumn{1}{r!{\huxvb{0, 0, 0}{0}}}{\huxtpad{0pt + 1em}\raggedleft \hspace{0pt} \hphantom{0}\hphantom{0}\hphantom{0} \hspace{0pt}\huxbpad{0pt}} &
\multicolumn{1}{r!{\huxvb{0, 0, 0}{0}}}{\huxtpad{0pt + 1em}\raggedleft \hspace{0pt} \hphantom{0}\hphantom{0}\hphantom{0} \hspace{0pt}\huxbpad{0pt}} &
\multicolumn{1}{r!{\huxvb{0, 0, 0}{0}}}{\huxtpad{0pt + 1em}\raggedleft \hspace{0pt} -0.34 \hspace{0pt}\huxbpad{0pt}} &
\multicolumn{1}{r!{\huxvb{0, 0, 0}{0}}}{\huxtpad{0pt + 1em}\raggedleft \hspace{0pt} -2.75 \hspace{0pt}\huxbpad{0pt}} &
\multicolumn{1}{r!{\huxvb{0, 0, 0}{0}}}{\huxtpad{0pt + 1em}\raggedleft \hspace{0pt} -0.40 \hspace{0pt}\huxbpad{0pt}} &
\multicolumn{1}{r!{\huxvb{0, 0, 0}{0}}}{\huxtpad{0pt + 1em}\raggedleft \hspace{0pt} -2.93 \hspace{0pt}\huxbpad{0pt}} \tabularnewline[-0.5pt]

\hhline{}
\arrayrulecolor{black}

\multicolumn{1}{!{\huxvb{0, 0, 0}{0}}l!{\huxvb{0, 0, 0}{0}}}{\huxtpad{0pt + 1em}\raggedright \hspace{0pt} RMW \hspace{0pt}\huxbpad{0pt}} &
\multicolumn{1}{r!{\huxvb{0, 0, 0}{0}}}{\huxtpad{0pt + 1em}\raggedleft \hspace{0pt} -0.32 \hspace{0pt}\huxbpad{0pt}} &
\multicolumn{1}{r!{\huxvb{0, 0, 0}{0}}}{\huxtpad{0pt + 1em}\raggedleft \hspace{0pt} -1.72 \hspace{0pt}\huxbpad{0pt}} &
\multicolumn{1}{r!{\huxvb{0, 0, 0}{0}}}{\huxtpad{0pt + 1em}\raggedleft \hspace{0pt} \hphantom{0}\hphantom{0}\hphantom{0} \hspace{0pt}\huxbpad{0pt}} &
\multicolumn{1}{r!{\huxvb{0, 0, 0}{0}}}{\huxtpad{0pt + 1em}\raggedleft \hspace{0pt} \hphantom{0}\hphantom{0}\hphantom{0} \hspace{0pt}\huxbpad{0pt}} &
\multicolumn{1}{r!{\huxvb{0, 0, 0}{0}}}{\huxtpad{0pt + 1em}\raggedleft \hspace{0pt} -0.34 \hspace{0pt}\huxbpad{0pt}} &
\multicolumn{1}{r!{\huxvb{0, 0, 0}{0}}}{\huxtpad{0pt + 1em}\raggedleft \hspace{0pt} -1.76 \hspace{0pt}\huxbpad{0pt}} &
\multicolumn{1}{r!{\huxvb{0, 0, 0}{0}}}{\huxtpad{0pt + 1em}\raggedleft \hspace{0pt} \hphantom{0}\hphantom{0}\hphantom{0} \hspace{0pt}\huxbpad{0pt}} &
\multicolumn{1}{r!{\huxvb{0, 0, 0}{0}}}{\huxtpad{0pt + 1em}\raggedleft \hspace{0pt} \hphantom{0}\hphantom{0}\hphantom{0} \hspace{0pt}\huxbpad{0pt}} \tabularnewline[-0.5pt]

\hhline{}
\arrayrulecolor{black}

\multicolumn{1}{!{\huxvb{0, 0, 0}{0}}l!{\huxvb{0, 0, 0}{0}}}{\huxtpad{0pt + 1em}\raggedright \hspace{0pt} CMA \hspace{0pt}\huxbpad{0pt}} &
\multicolumn{1}{r!{\huxvb{0, 0, 0}{0}}}{\huxtpad{0pt + 1em}\raggedleft \hspace{0pt} -0.46 \hspace{0pt}\huxbpad{0pt}} &
\multicolumn{1}{r!{\huxvb{0, 0, 0}{0}}}{\huxtpad{0pt + 1em}\raggedleft \hspace{0pt} -0.93 \hspace{0pt}\huxbpad{0pt}} &
\multicolumn{1}{r!{\huxvb{0, 0, 0}{0}}}{\huxtpad{0pt + 1em}\raggedleft \hspace{0pt} \hphantom{0}\hphantom{0}\hphantom{0} \hspace{0pt}\huxbpad{0pt}} &
\multicolumn{1}{r!{\huxvb{0, 0, 0}{0}}}{\huxtpad{0pt + 1em}\raggedleft \hspace{0pt} \hphantom{0}\hphantom{0}\hphantom{0} \hspace{0pt}\huxbpad{0pt}} &
\multicolumn{1}{r!{\huxvb{0, 0, 0}{0}}}{\huxtpad{0pt + 1em}\raggedleft \hspace{0pt} -0.05 \hspace{0pt}\huxbpad{0pt}} &
\multicolumn{1}{r!{\huxvb{0, 0, 0}{0}}}{\huxtpad{0pt + 1em}\raggedleft \hspace{0pt} -0.23 \hspace{0pt}\huxbpad{0pt}} &
\multicolumn{1}{r!{\huxvb{0, 0, 0}{0}}}{\huxtpad{0pt + 1em}\raggedleft \hspace{0pt} \hphantom{0}\hphantom{0}\hphantom{0} \hspace{0pt}\huxbpad{0pt}} &
\multicolumn{1}{r!{\huxvb{0, 0, 0}{0}}}{\huxtpad{0pt + 1em}\raggedleft \hspace{0pt} \hphantom{0}\hphantom{0}\hphantom{0} \hspace{0pt}\huxbpad{0pt}} \tabularnewline[-0.5pt]

\hhline{}
\arrayrulecolor{black}

\multicolumn{1}{!{\huxvb{0, 0, 0}{0}}l!{\huxvb{0, 0, 0}{0}}}{\huxtpad{0pt + 1em}\raggedright \hspace{0pt} R\_IA \hspace{0pt}\huxbpad{0pt}} &
\multicolumn{1}{r!{\huxvb{0, 0, 0}{0}}}{\huxtpad{0pt + 1em}\raggedleft \hspace{0pt} 0.78 \hspace{0pt}\huxbpad{0pt}} &
\multicolumn{1}{r!{\huxvb{0, 0, 0}{0}}}{\huxtpad{0pt + 1em}\raggedleft \hspace{0pt} 1.73 \hspace{0pt}\huxbpad{0pt}} &
\multicolumn{1}{r!{\huxvb{0, 0, 0}{0}}}{\huxtpad{0pt + 1em}\raggedleft \hspace{0pt} \hphantom{0}\hphantom{0}\hphantom{0} \hspace{0pt}\huxbpad{0pt}} &
\multicolumn{1}{r!{\huxvb{0, 0, 0}{0}}}{\huxtpad{0pt + 1em}\raggedleft \hspace{0pt} \hphantom{0}\hphantom{0}\hphantom{0} \hspace{0pt}\huxbpad{0pt}} &
\multicolumn{1}{r!{\huxvb{0, 0, 0}{0}}}{\huxtpad{0pt + 1em}\raggedleft \hspace{0pt} \hphantom{0}\hphantom{0}\hphantom{0} \hspace{0pt}\huxbpad{0pt}} &
\multicolumn{1}{r!{\huxvb{0, 0, 0}{0}}}{\huxtpad{0pt + 1em}\raggedleft \hspace{0pt} \hphantom{0}\hphantom{0}\hphantom{0} \hspace{0pt}\huxbpad{0pt}} &
\multicolumn{1}{r!{\huxvb{0, 0, 0}{0}}}{\huxtpad{0pt + 1em}\raggedleft \hspace{0pt} 0.03 \hspace{0pt}\huxbpad{0pt}} &
\multicolumn{1}{r!{\huxvb{0, 0, 0}{0}}}{\huxtpad{0pt + 1em}\raggedleft \hspace{0pt} 0.16 \hspace{0pt}\huxbpad{0pt}} \tabularnewline[-0.5pt]

\hhline{}
\arrayrulecolor{black}

\multicolumn{1}{!{\huxvb{0, 0, 0}{0}}l!{\huxvb{0, 0, 0}{0}}}{\huxtpad{0pt + 1em}\raggedright \hspace{0pt} R\_ROE \hspace{0pt}\huxbpad{0pt}} &
\multicolumn{1}{r!{\huxvb{0, 0, 0}{0}}}{\huxtpad{0pt + 1em}\raggedleft \hspace{0pt} 0.02 \hspace{0pt}\huxbpad{0pt}} &
\multicolumn{1}{r!{\huxvb{0, 0, 0}{0}}}{\huxtpad{0pt + 1em}\raggedleft \hspace{0pt} 0.11 \hspace{0pt}\huxbpad{0pt}} &
\multicolumn{1}{r!{\huxvb{0, 0, 0}{0}}}{\huxtpad{0pt + 1em}\raggedleft \hspace{0pt} \hphantom{0}\hphantom{0}\hphantom{0} \hspace{0pt}\huxbpad{0pt}} &
\multicolumn{1}{r!{\huxvb{0, 0, 0}{0}}}{\huxtpad{0pt + 1em}\raggedleft \hspace{0pt} \hphantom{0}\hphantom{0}\hphantom{0} \hspace{0pt}\huxbpad{0pt}} &
\multicolumn{1}{r!{\huxvb{0, 0, 0}{0}}}{\huxtpad{0pt + 1em}\raggedleft \hspace{0pt} \hphantom{0}\hphantom{0}\hphantom{0} \hspace{0pt}\huxbpad{0pt}} &
\multicolumn{1}{r!{\huxvb{0, 0, 0}{0}}}{\huxtpad{0pt + 1em}\raggedleft \hspace{0pt} \hphantom{0}\hphantom{0}\hphantom{0} \hspace{0pt}\huxbpad{0pt}} &
\multicolumn{1}{r!{\huxvb{0, 0, 0}{0}}}{\huxtpad{0pt + 1em}\raggedleft \hspace{0pt} -0.31 \hspace{0pt}\huxbpad{0pt}} &
\multicolumn{1}{r!{\huxvb{0, 0, 0}{0}}}{\huxtpad{0pt + 1em}\raggedleft \hspace{0pt} -1.25 \hspace{0pt}\huxbpad{0pt}} \tabularnewline[-0.5pt]

\hhline{}
\arrayrulecolor{black}

\multicolumn{1}{!{\huxvb{0, 0, 0}{0}}l!{\huxvb{0, 0, 0}{0}}}{\huxtpad{0pt + 1em}\raggedright \hspace{0pt} R\_EG \hspace{0pt}\huxbpad{0pt}} &
\multicolumn{1}{r!{\huxvb{0, 0, 0}{0}}}{\huxtpad{0pt + 1em}\raggedleft \hspace{0pt} 0.02 \hspace{0pt}\huxbpad{0pt}} &
\multicolumn{1}{r!{\huxvb{0, 0, 0}{0}}}{\huxtpad{0pt + 1em}\raggedleft \hspace{0pt} 0.10 \hspace{0pt}\huxbpad{0pt}} &
\multicolumn{1}{r!{\huxvb{0, 0, 0}{0}}}{\huxtpad{0pt + 1em}\raggedleft \hspace{0pt} \hphantom{0}\hphantom{0}\hphantom{0} \hspace{0pt}\huxbpad{0pt}} &
\multicolumn{1}{r!{\huxvb{0, 0, 0}{0}}}{\huxtpad{0pt + 1em}\raggedleft \hspace{0pt} \hphantom{0}\hphantom{0}\hphantom{0} \hspace{0pt}\huxbpad{0pt}} &
\multicolumn{1}{r!{\huxvb{0, 0, 0}{0}}}{\huxtpad{0pt + 1em}\raggedleft \hspace{0pt} \hphantom{0}\hphantom{0}\hphantom{0} \hspace{0pt}\huxbpad{0pt}} &
\multicolumn{1}{r!{\huxvb{0, 0, 0}{0}}}{\huxtpad{0pt + 1em}\raggedleft \hspace{0pt} \hphantom{0}\hphantom{0}\hphantom{0} \hspace{0pt}\huxbpad{0pt}} &
\multicolumn{1}{r!{\huxvb{0, 0, 0}{0}}}{\huxtpad{0pt + 1em}\raggedleft \hspace{0pt} -0.01 \hspace{0pt}\huxbpad{0pt}} &
\multicolumn{1}{r!{\huxvb{0, 0, 0}{0}}}{\huxtpad{0pt + 1em}\raggedleft \hspace{0pt} -0.05 \hspace{0pt}\huxbpad{0pt}} \tabularnewline[-0.5pt]

\hhline{}
\arrayrulecolor{black}

\multicolumn{1}{!{\huxvb{0, 0, 0}{0}}l!{\huxvb{0, 0, 0}{0}}}{\huxtpad{0pt + 1em}\raggedright \hspace{0pt} $R^2$ \hspace{0pt}\huxbpad{0pt}} &
\multicolumn{1}{r!{\huxvb{0, 0, 0}{0}}}{\huxtpad{0pt + 1em}\raggedleft \hspace{0pt} 12.82 \hspace{0pt}\huxbpad{0pt}} &
\multicolumn{1}{r!{\huxvb{0, 0, 0}{0}}}{\huxtpad{0pt + 1em}\raggedleft \hspace{0pt} \hphantom{0}\hphantom{0}\hphantom{0} \hspace{0pt}\huxbpad{0pt}} &
\multicolumn{1}{r!{\huxvb{0, 0, 0}{0}}}{\huxtpad{0pt + 1em}\raggedleft \hspace{0pt} 9.65 \hspace{0pt}\huxbpad{0pt}} &
\multicolumn{1}{r!{\huxvb{0, 0, 0}{0}}}{\huxtpad{0pt + 1em}\raggedleft \hspace{0pt} \hphantom{0}\hphantom{0}\hphantom{0} \hspace{0pt}\huxbpad{0pt}} &
\multicolumn{1}{r!{\huxvb{0, 0, 0}{0}}}{\huxtpad{0pt + 1em}\raggedleft \hspace{0pt} 4.94 \hspace{0pt}\huxbpad{0pt}} &
\multicolumn{1}{r!{\huxvb{0, 0, 0}{0}}}{\huxtpad{0pt + 1em}\raggedleft \hspace{0pt} \hphantom{0}\hphantom{0}\hphantom{0} \hspace{0pt}\huxbpad{0pt}} &
\multicolumn{1}{r!{\huxvb{0, 0, 0}{0}}}{\huxtpad{0pt + 1em}\raggedleft \hspace{0pt} 4.72 \hspace{0pt}\huxbpad{0pt}} &
\multicolumn{1}{r!{\huxvb{0, 0, 0}{0}}}{\huxtpad{0pt + 1em}\raggedleft \hspace{0pt} \hphantom{0}\hphantom{0}\hphantom{0} \hspace{0pt}\huxbpad{0pt}} \tabularnewline[-0.5pt]

\hhline{>{\huxb{0, 0, 0}{0.8}}->{\huxb{0, 0, 0}{0.8}}->{\huxb{0, 0, 0}{0.8}}->{\huxb{0, 0, 0}{0.8}}->{\huxb{0, 0, 0}{0.8}}->{\huxb{0, 0, 0}{0.8}}->{\huxb{0, 0, 0}{0.8}}->{\huxb{0, 0, 0}{0.8}}->{\huxb{0, 0, 0}{0.8}}-}
\arrayrulecolor{black}
\end{tabular}
    \end{adjustbox}

    \vspace{5pt}
    Panel B: StockGPT Spans Stock Factors
    \vspace{5pt}
    
    \begin{adjustbox}{width = \columnwidth, center}
        
  \providecommand{\huxb}[2]{\arrayrulecolor[RGB]{#1}\global\arrayrulewidth=#2pt}
  \providecommand{\huxvb}[2]{\color[RGB]{#1}\vrule width #2pt}
  \providecommand{\huxtpad}[1]{\rule{0pt}{#1}}
  \providecommand{\huxbpad}[1]{\rule[-#1]{0pt}{#1}}
\begin{tabular}{l l l l l l l l l l l l}

\hhline{>{\huxb{0, 0, 0}{0.8}}->{\huxb{0, 0, 0}{0.8}}->{\huxb{0, 0, 0}{0.8}}->{\huxb{0, 0, 0}{0.8}}->{\huxb{0, 0, 0}{0.8}}->{\huxb{0, 0, 0}{0.8}}->{\huxb{0, 0, 0}{0.8}}->{\huxb{0, 0, 0}{0.8}}->{\huxb{0, 0, 0}{0.8}}->{\huxb{0, 0, 0}{0.8}}->{\huxb{0, 0, 0}{0.8}}->{\huxb{0, 0, 0}{0.8}}-}
\arrayrulecolor{black}

\multicolumn{1}{!{\huxvb{0, 0, 0}{0}}l!{\huxvb{0, 0, 0}{0}}}{\huxtpad{0pt + 1em}\raggedright \hspace{0pt}  \hspace{0pt}\huxbpad{0pt}} &
\multicolumn{1}{r!{\huxvb{0, 0, 0}{0}}}{\huxtpad{0pt + 1em}\raggedleft \hspace{0pt} ST\_Rev \hspace{0pt}\huxbpad{0pt}} &
\multicolumn{1}{r!{\huxvb{0, 0, 0}{0}}}{\huxtpad{0pt + 1em}\raggedleft \hspace{0pt} Mom \hspace{0pt}\huxbpad{0pt}} &
\multicolumn{1}{r!{\huxvb{0, 0, 0}{0}}}{\huxtpad{0pt + 1em}\raggedleft \hspace{0pt} LT\_Rev \hspace{0pt}\huxbpad{0pt}} &
\multicolumn{1}{r!{\huxvb{0, 0, 0}{0}}}{\huxtpad{0pt + 1em}\raggedleft \hspace{0pt} MKT \hspace{0pt}\huxbpad{0pt}} &
\multicolumn{1}{r!{\huxvb{0, 0, 0}{0}}}{\huxtpad{0pt + 1em}\raggedleft \hspace{0pt} HML \hspace{0pt}\huxbpad{0pt}} &
\multicolumn{1}{r!{\huxvb{0, 0, 0}{0}}}{\huxtpad{0pt + 1em}\raggedleft \hspace{0pt} SMB \hspace{0pt}\huxbpad{0pt}} &
\multicolumn{1}{r!{\huxvb{0, 0, 0}{0}}}{\huxtpad{0pt + 1em}\raggedleft \hspace{0pt} RMW \hspace{0pt}\huxbpad{0pt}} &
\multicolumn{1}{r!{\huxvb{0, 0, 0}{0}}}{\huxtpad{0pt + 1em}\raggedleft \hspace{0pt} CMA \hspace{0pt}\huxbpad{0pt}} &
\multicolumn{1}{r!{\huxvb{0, 0, 0}{0}}}{\huxtpad{0pt + 1em}\raggedleft \hspace{0pt} R\_IA \hspace{0pt}\huxbpad{0pt}} &
\multicolumn{1}{r!{\huxvb{0, 0, 0}{0}}}{\huxtpad{0pt + 1em}\raggedleft \hspace{0pt} R\_ROE \hspace{0pt}\huxbpad{0pt}} &
\multicolumn{1}{r!{\huxvb{0, 0, 0}{0}}}{\huxtpad{0pt + 1em}\raggedleft \hspace{0pt} R\_EG \hspace{0pt}\huxbpad{0pt}} \tabularnewline[-0.5pt]

\hhline{>{\huxb{0, 0, 0}{0.8}}->{\huxb{0, 0, 0}{0.8}}->{\huxb{0, 0, 0}{0.8}}->{\huxb{0, 0, 0}{0.8}}->{\huxb{0, 0, 0}{0.8}}->{\huxb{0, 0, 0}{0.8}}->{\huxb{0, 0, 0}{0.8}}->{\huxb{0, 0, 0}{0.8}}->{\huxb{0, 0, 0}{0.8}}->{\huxb{0, 0, 0}{0.8}}->{\huxb{0, 0, 0}{0.8}}->{\huxb{0, 0, 0}{0.8}}-}
\arrayrulecolor{black}

\multicolumn{1}{!{\huxvb{0, 0, 0}{0}}l!{\huxvb{0, 0, 0}{0}}}{\huxtpad{0pt + 1em}\raggedright \hspace{0pt} $\alpha$ \hspace{0pt}\huxbpad{0pt}} &
\multicolumn{1}{r!{\huxvb{0, 0, 0}{0}}}{\huxtpad{0pt + 1em}\raggedleft \hspace{0pt} 9.00 \hspace{0pt}\huxbpad{0pt}} &
\multicolumn{1}{r!{\huxvb{0, 0, 0}{0}}}{\huxtpad{0pt + 1em}\raggedleft \hspace{0pt} 5.83 \hspace{0pt}\huxbpad{0pt}} &
\multicolumn{1}{r!{\huxvb{0, 0, 0}{0}}}{\huxtpad{0pt + 1em}\raggedleft \hspace{0pt} 2.33 \hspace{0pt}\huxbpad{0pt}} &
\multicolumn{1}{r!{\huxvb{0, 0, 0}{0}}}{\huxtpad{0pt + 1em}\raggedleft \hspace{0pt} 6.15 \hspace{0pt}\huxbpad{0pt}} &
\multicolumn{1}{r!{\huxvb{0, 0, 0}{0}}}{\huxtpad{0pt + 1em}\raggedleft \hspace{0pt} 2.14 \hspace{0pt}\huxbpad{0pt}} &
\multicolumn{1}{r!{\huxvb{0, 0, 0}{0}}}{\huxtpad{0pt + 1em}\raggedleft \hspace{0pt} 3.35 \hspace{0pt}\huxbpad{0pt}} &
\multicolumn{1}{r!{\huxvb{0, 0, 0}{0}}}{\huxtpad{0pt + 1em}\raggedleft \hspace{0pt} 6.06 \hspace{0pt}\huxbpad{0pt}} &
\multicolumn{1}{r!{\huxvb{0, 0, 0}{0}}}{\huxtpad{0pt + 1em}\raggedleft \hspace{0pt} 2.84 \hspace{0pt}\huxbpad{0pt}} &
\multicolumn{1}{r!{\huxvb{0, 0, 0}{0}}}{\huxtpad{0pt + 1em}\raggedleft \hspace{0pt} 2.26 \hspace{0pt}\huxbpad{0pt}} &
\multicolumn{1}{r!{\huxvb{0, 0, 0}{0}}}{\huxtpad{0pt + 1em}\raggedleft \hspace{0pt} 5.44 \hspace{0pt}\huxbpad{0pt}} &
\multicolumn{1}{r!{\huxvb{0, 0, 0}{0}}}{\huxtpad{0pt + 1em}\raggedleft \hspace{0pt} 5.69 \hspace{0pt}\huxbpad{0pt}} \tabularnewline[-0.5pt]

\hhline{}
\arrayrulecolor{black}

\multicolumn{1}{!{\huxvb{0, 0, 0}{0}}l!{\huxvb{0, 0, 0}{0}}}{\huxtpad{0pt + 1em}\raggedright \hspace{0pt} $t_{\alpha}$ \hspace{0pt}\huxbpad{0pt}} &
\multicolumn{1}{r!{\huxvb{0, 0, 0}{0}}}{\huxtpad{0pt + 1em}\raggedleft \hspace{0pt} 4.03 \hspace{0pt}\huxbpad{0pt}} &
\multicolumn{1}{r!{\huxvb{0, 0, 0}{0}}}{\huxtpad{0pt + 1em}\raggedleft \hspace{0pt} 1.98 \hspace{0pt}\huxbpad{0pt}} &
\multicolumn{1}{r!{\huxvb{0, 0, 0}{0}}}{\huxtpad{0pt + 1em}\raggedleft \hspace{0pt} 0.97 \hspace{0pt}\huxbpad{0pt}} &
\multicolumn{1}{r!{\huxvb{0, 0, 0}{0}}}{\huxtpad{0pt + 1em}\raggedleft \hspace{0pt} 1.65 \hspace{0pt}\huxbpad{0pt}} &
\multicolumn{1}{r!{\huxvb{0, 0, 0}{0}}}{\huxtpad{0pt + 1em}\raggedleft \hspace{0pt} 0.74 \hspace{0pt}\huxbpad{0pt}} &
\multicolumn{1}{r!{\huxvb{0, 0, 0}{0}}}{\huxtpad{0pt + 1em}\raggedleft \hspace{0pt} 1.91 \hspace{0pt}\huxbpad{0pt}} &
\multicolumn{1}{r!{\huxvb{0, 0, 0}{0}}}{\huxtpad{0pt + 1em}\raggedleft \hspace{0pt} 3.10 \hspace{0pt}\huxbpad{0pt}} &
\multicolumn{1}{r!{\huxvb{0, 0, 0}{0}}}{\huxtpad{0pt + 1em}\raggedleft \hspace{0pt} 1.50 \hspace{0pt}\huxbpad{0pt}} &
\multicolumn{1}{r!{\huxvb{0, 0, 0}{0}}}{\huxtpad{0pt + 1em}\raggedleft \hspace{0pt} 1.11 \hspace{0pt}\huxbpad{0pt}} &
\multicolumn{1}{r!{\huxvb{0, 0, 0}{0}}}{\huxtpad{0pt + 1em}\raggedleft \hspace{0pt} 2.77 \hspace{0pt}\huxbpad{0pt}} &
\multicolumn{1}{r!{\huxvb{0, 0, 0}{0}}}{\huxtpad{0pt + 1em}\raggedleft \hspace{0pt} 2.95 \hspace{0pt}\huxbpad{0pt}} \tabularnewline[-0.5pt]

\hhline{}
\arrayrulecolor{black}

\multicolumn{1}{!{\huxvb{0, 0, 0}{0}}l!{\huxvb{0, 0, 0}{0}}}{\huxtpad{0pt + 1em}\raggedright \hspace{0pt} $\beta$ \hspace{0pt}\huxbpad{0pt}} &
\multicolumn{1}{r!{\huxvb{0, 0, 0}{0}}}{\huxtpad{0pt + 1em}\raggedleft \hspace{0pt} -0.02 \hspace{0pt}\huxbpad{0pt}} &
\multicolumn{1}{r!{\huxvb{0, 0, 0}{0}}}{\huxtpad{0pt + 1em}\raggedleft \hspace{0pt} -0.26 \hspace{0pt}\huxbpad{0pt}} &
\multicolumn{1}{r!{\huxvb{0, 0, 0}{0}}}{\huxtpad{0pt + 1em}\raggedleft \hspace{0pt} -0.14 \hspace{0pt}\huxbpad{0pt}} &
\multicolumn{1}{r!{\huxvb{0, 0, 0}{0}}}{\huxtpad{0pt + 1em}\raggedleft \hspace{0pt} 0.10 \hspace{0pt}\huxbpad{0pt}} &
\multicolumn{1}{r!{\huxvb{0, 0, 0}{0}}}{\huxtpad{0pt + 1em}\raggedleft \hspace{0pt} -0.07 \hspace{0pt}\huxbpad{0pt}} &
\multicolumn{1}{r!{\huxvb{0, 0, 0}{0}}}{\huxtpad{0pt + 1em}\raggedleft \hspace{0pt} -0.08 \hspace{0pt}\huxbpad{0pt}} &
\multicolumn{1}{r!{\huxvb{0, 0, 0}{0}}}{\huxtpad{0pt + 1em}\raggedleft \hspace{0pt} -0.08 \hspace{0pt}\huxbpad{0pt}} &
\multicolumn{1}{r!{\huxvb{0, 0, 0}{0}}}{\huxtpad{0pt + 1em}\raggedleft \hspace{0pt} -0.04 \hspace{0pt}\huxbpad{0pt}} &
\multicolumn{1}{r!{\huxvb{0, 0, 0}{0}}}{\huxtpad{0pt + 1em}\raggedleft \hspace{0pt} -0.02 \hspace{0pt}\huxbpad{0pt}} &
\multicolumn{1}{r!{\huxvb{0, 0, 0}{0}}}{\huxtpad{0pt + 1em}\raggedleft \hspace{0pt} -0.09 \hspace{0pt}\huxbpad{0pt}} &
\multicolumn{1}{r!{\huxvb{0, 0, 0}{0}}}{\huxtpad{0pt + 1em}\raggedleft \hspace{0pt} -0.03 \hspace{0pt}\huxbpad{0pt}} \tabularnewline[-0.5pt]

\hhline{}
\arrayrulecolor{black}

\multicolumn{1}{!{\huxvb{0, 0, 0}{0}}l!{\huxvb{0, 0, 0}{0}}}{\huxtpad{0pt + 1em}\raggedright \hspace{0pt} $t_{\beta}$ \hspace{0pt}\huxbpad{0pt}} &
\multicolumn{1}{r!{\huxvb{0, 0, 0}{0}}}{\huxtpad{0pt + 1em}\raggedleft \hspace{0pt} -0.21 \hspace{0pt}\huxbpad{0pt}} &
\multicolumn{1}{r!{\huxvb{0, 0, 0}{0}}}{\huxtpad{0pt + 1em}\raggedleft \hspace{0pt} -1.60 \hspace{0pt}\huxbpad{0pt}} &
\multicolumn{1}{r!{\huxvb{0, 0, 0}{0}}}{\huxtpad{0pt + 1em}\raggedleft \hspace{0pt} -2.03 \hspace{0pt}\huxbpad{0pt}} &
\multicolumn{1}{r!{\huxvb{0, 0, 0}{0}}}{\huxtpad{0pt + 1em}\raggedleft \hspace{0pt} 1.04 \hspace{0pt}\huxbpad{0pt}} &
\multicolumn{1}{r!{\huxvb{0, 0, 0}{0}}}{\huxtpad{0pt + 1em}\raggedleft \hspace{0pt} -0.78 \hspace{0pt}\huxbpad{0pt}} &
\multicolumn{1}{r!{\huxvb{0, 0, 0}{0}}}{\huxtpad{0pt + 1em}\raggedleft \hspace{0pt} -1.40 \hspace{0pt}\huxbpad{0pt}} &
\multicolumn{1}{r!{\huxvb{0, 0, 0}{0}}}{\huxtpad{0pt + 1em}\raggedleft \hspace{0pt} -1.39 \hspace{0pt}\huxbpad{0pt}} &
\multicolumn{1}{r!{\huxvb{0, 0, 0}{0}}}{\huxtpad{0pt + 1em}\raggedleft \hspace{0pt} -0.77 \hspace{0pt}\huxbpad{0pt}} &
\multicolumn{1}{r!{\huxvb{0, 0, 0}{0}}}{\huxtpad{0pt + 1em}\raggedleft \hspace{0pt} -0.33 \hspace{0pt}\huxbpad{0pt}} &
\multicolumn{1}{r!{\huxvb{0, 0, 0}{0}}}{\huxtpad{0pt + 1em}\raggedleft \hspace{0pt} -0.99 \hspace{0pt}\huxbpad{0pt}} &
\multicolumn{1}{r!{\huxvb{0, 0, 0}{0}}}{\huxtpad{0pt + 1em}\raggedleft \hspace{0pt} -0.46 \hspace{0pt}\huxbpad{0pt}} \tabularnewline[-0.5pt]

\hhline{}
\arrayrulecolor{black}

\multicolumn{1}{!{\huxvb{0, 0, 0}{0}}l!{\huxvb{0, 0, 0}{0}}}{\huxtpad{0pt + 1em}\raggedright \hspace{0pt} $R^2$ \hspace{0pt}\huxbpad{0pt}} &
\multicolumn{1}{r!{\huxvb{0, 0, 0}{0}}}{\huxtpad{0pt + 1em}\raggedleft \hspace{0pt} -0.33 \hspace{0pt}\huxbpad{0pt}} &
\multicolumn{1}{r!{\huxvb{0, 0, 0}{0}}}{\huxtpad{0pt + 1em}\raggedleft \hspace{0pt} 5.05 \hspace{0pt}\huxbpad{0pt}} &
\multicolumn{1}{r!{\huxvb{0, 0, 0}{0}}}{\huxtpad{0pt + 1em}\raggedleft \hspace{0pt} 3.46 \hspace{0pt}\huxbpad{0pt}} &
\multicolumn{1}{r!{\huxvb{0, 0, 0}{0}}}{\huxtpad{0pt + 1em}\raggedleft \hspace{0pt} 0.53 \hspace{0pt}\huxbpad{0pt}} &
\multicolumn{1}{r!{\huxvb{0, 0, 0}{0}}}{\huxtpad{0pt + 1em}\raggedleft \hspace{0pt} 0.43 \hspace{0pt}\huxbpad{0pt}} &
\multicolumn{1}{r!{\huxvb{0, 0, 0}{0}}}{\huxtpad{0pt + 1em}\raggedleft \hspace{0pt} 1.34 \hspace{0pt}\huxbpad{0pt}} &
\multicolumn{1}{r!{\huxvb{0, 0, 0}{0}}}{\huxtpad{0pt + 1em}\raggedleft \hspace{0pt} 2.08 \hspace{0pt}\huxbpad{0pt}} &
\multicolumn{1}{r!{\huxvb{0, 0, 0}{0}}}{\huxtpad{0pt + 1em}\raggedleft \hspace{0pt} 0.36 \hspace{0pt}\huxbpad{0pt}} &
\multicolumn{1}{r!{\huxvb{0, 0, 0}{0}}}{\huxtpad{0pt + 1em}\raggedleft \hspace{0pt} -0.26 \hspace{0pt}\huxbpad{0pt}} &
\multicolumn{1}{r!{\huxvb{0, 0, 0}{0}}}{\huxtpad{0pt + 1em}\raggedleft \hspace{0pt} 1.28 \hspace{0pt}\huxbpad{0pt}} &
\multicolumn{1}{r!{\huxvb{0, 0, 0}{0}}}{\huxtpad{0pt + 1em}\raggedleft \hspace{0pt} -0.17 \hspace{0pt}\huxbpad{0pt}} \tabularnewline[-0.5pt]

\hhline{>{\huxb{0, 0, 0}{0.8}}->{\huxb{0, 0, 0}{0.8}}->{\huxb{0, 0, 0}{0.8}}->{\huxb{0, 0, 0}{0.8}}->{\huxb{0, 0, 0}{0.8}}->{\huxb{0, 0, 0}{0.8}}->{\huxb{0, 0, 0}{0.8}}->{\huxb{0, 0, 0}{0.8}}->{\huxb{0, 0, 0}{0.8}}->{\huxb{0, 0, 0}{0.8}}->{\huxb{0, 0, 0}{0.8}}->{\huxb{0, 0, 0}{0.8}}-}
\arrayrulecolor{black}
\end{tabular}
    \end{adjustbox}
    
    \label{tab:monthly_alpha}
\end{table}

\clearpage

\begin{appendices}
\section{Daily Model Trained will All Stocks}
\begin{table}[ht]
    \caption{Daily Portfolio Statistics}

    \vspace{5pt}
    \footnotesize
    \noindent This table reports the return statistics of the daily long-short StockGPT-based portfolios. Mean and SD (standard deviation) are in annualized percentage points; Mean/SD (Sharpe ratio) is annualized; Min, Max, and MDD (max drawdown) are in percentage points; and t-Mean is $t$-statistic of the mean portfolio return using Newey-West standard error with 20 lags. Portfolios are formed after excluding stocks in the bottom decile based on market value. Horizon 1 (2) refers to using return forecasts for day $t+1$ to form portfolios for day $t+1$ ($t+2$). EW (VW) refers to equal-weighting (value weighting). Price Filter refers to the price level under which stocks are removed. The sample is daily from January 2001 to December 2023.
    \vspace{5pt}
    
    \centering
    \begin{adjustbox}{width = \columnwidth, center}
        
  \providecommand{\huxb}[2]{\arrayrulecolor[RGB]{#1}\global\arrayrulewidth=#2pt}
  \providecommand{\huxvb}[2]{\color[RGB]{#1}\vrule width #2pt}
  \providecommand{\huxtpad}[1]{\rule{0pt}{#1}}
  \providecommand{\huxbpad}[1]{\rule[-#1]{0pt}{#1}}
\begin{tabular}{l l l l l l l l l l}

\hhline{>{\huxb{0, 0, 0}{0.8}}->{\huxb{0, 0, 0}{0.8}}->{\huxb{0, 0, 0}{0.8}}->{\huxb{0, 0, 0}{0.8}}->{\huxb{0, 0, 0}{0.8}}->{\huxb{0, 0, 0}{0.8}}->{\huxb{0, 0, 0}{0.8}}->{\huxb{0, 0, 0}{0.8}}->{\huxb{0, 0, 0}{0.8}}->{\huxb{0, 0, 0}{0.8}}-}
\arrayrulecolor{black}

\multicolumn{1}{!{\huxvb{0, 0, 0}{0}}c!{\huxvb{0, 0, 0}{0}}}{\huxtpad{0pt + 1em}\centering \hspace{0pt} Horizon \hspace{0pt}\huxbpad{0pt}} &
\multicolumn{1}{c!{\huxvb{0, 0, 0}{0}}}{\huxtpad{0pt + 1em}\centering \hspace{0pt} Weight \hspace{0pt}\huxbpad{0pt}} &
\multicolumn{1}{c!{\huxvb{0, 0, 0}{0}}}{\huxtpad{0pt + 1em}\centering \hspace{0pt} Price Filter \hspace{0pt}\huxbpad{0pt}} &
\multicolumn{1}{c!{\huxvb{0, 0, 0}{0}}}{\huxtpad{0pt + 1em}\centering \hspace{0pt} Mean \hspace{0pt}\huxbpad{0pt}} &
\multicolumn{1}{c!{\huxvb{0, 0, 0}{0}}}{\huxtpad{0pt + 1em}\centering \hspace{0pt} t-Mean \hspace{0pt}\huxbpad{0pt}} &
\multicolumn{1}{c!{\huxvb{0, 0, 0}{0}}}{\huxtpad{0pt + 1em}\centering \hspace{0pt} SD \hspace{0pt}\huxbpad{0pt}} &
\multicolumn{1}{c!{\huxvb{0, 0, 0}{0}}}{\huxtpad{0pt + 1em}\centering \hspace{0pt} Mean/SD \hspace{0pt}\huxbpad{0pt}} &
\multicolumn{1}{c!{\huxvb{0, 0, 0}{0}}}{\huxtpad{0pt + 1em}\centering \hspace{0pt} Min \hspace{0pt}\huxbpad{0pt}} &
\multicolumn{1}{c!{\huxvb{0, 0, 0}{0}}}{\huxtpad{0pt + 1em}\centering \hspace{0pt} Max \hspace{0pt}\huxbpad{0pt}} &
\multicolumn{1}{c!{\huxvb{0, 0, 0}{0}}}{\huxtpad{0pt + 1em}\centering \hspace{0pt} MDD \hspace{0pt}\huxbpad{0pt}} \tabularnewline
\end{tabular}
    \end{adjustbox}
    
    \label{tab:daily_port_all}
\end{table}
\end{appendices}

\end{document}